\documentclass{article}
\usepackage{graphicx} 
\usepackage{natbib}
\usepackage[title]{appendix}
\usepackage{amsmath}
\usepackage{fullpage}
\usepackage{xcolor}
\usepackage{subcaption}
\usepackage{comment}
\usepackage{authblk}
\usepackage{tabularx}
\usepackage{url}
\usepackage[table]{xcolor} 

\begin{document}

\title{Modeling U.S. Mortality and Suicide Rates by Integrating Mental Health and Socio-Economic Indicators}
\author[a]{Brianne Weaver}
\author[a]{Brigg Trendler}
\author[b]{Chris Groendyke}
\author[a]{Brian Hartman\footnote{Corresponding author - email: hartman@stat.byu.edu}}
\author[a]{Robert Richardson}
\author[c]{Davey Erekson}

\affil[a]{Department of Statistics, Brigham Young University, Provo, UT, USA}
\affil[b]{Department of Mathematics, Robert Morris University, Moon Township, PA, USA}
\affil[c]{Counseling and Psychological Services, Brigham Young University, Provo, UT, USA}

\maketitle

\begin{abstract}
Accurate mortality modeling is central to actuarial science and public health, especially as mental health emerges as a significant factor in population outcomes. This paper develops and applies a Bayesian hierarchical model to analyze U.S. county-level mortality and suicide rates from 2010 to 2023. Applying a conditional autoregressive (CAR) structure to each combination of sex and age grouping, the model captures spatial and temporal trends while incorporating mental health surveillance data and socio-economic indicators. We first assess socio-economic covariates in predicting suicide. While the results vary considerably by age and sex, we find that the county-wide levels of educational attainment, housing prices, marriage rates, racial composition, household size, and poor mental health days all have significant relationships with suicide rates. We next consider the impact of various mental health indicators on all-cause and suicide-specific mortality and find that the strongest effects are observed in younger populations. The spatial and temporal correlation structures reveal substantial regional clustering and time-consistent trends in both all-cause mortality and suicide rates, supporting the use of spatio-temporal methods. Our findings highlight the value of integrating mental health surveillance data into mortality models to better identify emerging risk areas and vulnerable populations. This approach has the potential to inform public health policy, resource allocation, and targeted interventions aimed at reducing disparities in mortality and suicide across U.S. communities.
\end{abstract}

\section{Introduction}

There is a clear and obvious need for accurate models of human mortality, and for centuries actuaries have been working to improve the quality of these models. (See, for example, \citet{campbell‐kelly2003life} for a summary of the early history of mortality models and life tables.) These improvements have taken various forms, including everything from better data collection to more complex mathematical descriptions and analytical methods to the incorporation of additional types of data or covariates into the model.

The notion of accounting for spatial and/or temporal effects in the modeling of mortality is not new. Since at least as far back as \citet{lee1992modeling}, researchers have developed increasingly intricate methods of incorporating spatio-temporal effects into mortality models. Many of these improvements have resulted in significantly more robust and accurate actuarial analyses; see Section \ref{subsec:litreviewspatial} for an overview. However, the idea of explicitly incorporating data related to mental health (and other related socio-economic factors) into mortality models is more novel, with the bulk of this work having been done in the last decade (see Section \ref{subsec:litreviewmental}). And importantly, few papers have attempted to integrate all three of these factors (spatial trends, temporal trends, and mental health indicators) into  a unified mortality model. 

The availability of increasingly detailed mental health surveillance data, combined with advances in Bayesian computational methods, creates unprecedented opportunities for developing these integrated models. Our current work focuses on creating frameworks that can efficiently handle the complex hierarchical structure of mortality data while incorporating mental health indicators as both predictors and mediators of geographic mortality variation. Specifically, we use a conditional auto-regressive (CAR) Bayesian hierarchical model to describe mortality, allowing us to incorporate both spatial and temporal effects, as well as their interaction. We apply this model to county-level data in the continental U.S., using various mental health indicators to predict mortality rates generally and, more specifically, suicide rates. In a separate analysis, we use a number of socio-economic covariates to predict suicide rates.

Supplementing the spatio-temporal mortality model with mental health data in this manner has a number of important potential benefits. It not only gives us the ability to more accurately predict and model mortality and suicides, but also allows us to identify emerging trends and ``at-risk'' populations that may have otherwise gone undetected. This information can then be used to shape public policy and to help better allocate resources.

The remainder of the paper is organized as follows: Section \ref{sec:litreview} discusses some of the previous work done on the spatio-temporal modeling of mortality curves and how mental health and socio-economic variables can be related to mortality rates; Section \ref{sec:data} describes the mortality data, mental health data, and socio-economic covariates used in this study; Section \ref{sec:methods} explains the statistical models and assumptions utilized in the analysis; Section \ref{sec:results} presents and discusses our results; and finally, Section \ref{sec:conc} gives some concluding remarks.

\section{Literature Review}
\label{sec:litreview}

\subsection{Modeling Spatial and Temporal Trends in Mortality}
\label{subsec:litreviewspatial}

The foundation for modern mortality modeling was established by the Lee-Carter model, which decomposed mortality rates into age-specific components and time trends, providing a widely adopted framework for demographic forecasting \citep{lee1992modeling}. Many researchers have since sought to extend and improve upon this model; see \citet{booth2008mortality} or \citet{lee2000lee} for discussions of such methods. Building upon this foundation, researchers developed Bayesian hierarchical models that extended mortality modeling to incorporate spatial dimensions alongside temporal patterns. For example, \citet{alexander2017flexible} demonstrated how Bayesian approaches could estimate subnational mortality by pooling information across geographic space while smoothing over time, using principal components to capture age patterns. \citet{goicoa2019flexible} used splines to model and smooth age-space-time patterns of mortality, specifically studying breast cancer mortality in Spain. 

The spatial-temporal framework was further refined through models that treat both area-specific intercepts and trends as correlated random effects, allowing for explicit modeling of space-time variation in mortality risk \citep{bernardinelli1995bayesian}. These Bayesian spatio-temporal models typically incorporate structured priors that account for spatial correlation through conditional autoregressive specifications \citep{besag1991bayesian} or Gaussian process frameworks \citep{banerjee2003hierarchical}. \citet{saavedra2021bayesian} used a Bayesian spatio-temporal model in order to estimate excess deaths in Spain due to COVID-19. The computational challenges inherent in these complex hierarchical models have been addressed through the development of Integrated Nested Laplace Approximations (INLA), which offers significant advantages over traditional Markov chain Monte Carlo (MCMC) methods in terms of computational efficiency and implementation ease for Bayesian spatial modeling \citep{rue2017, lindgren2015}. Recent work has extended these approaches to multivariate spatio-temporal settings for mortality modeling, such as the county-level analysis of U.S. mortality by \citet{Shull2025}.

Recent methodological advances have focused on handling large spatial datasets through nearest-neighbor Gaussian process models \citep{datta2016hierarchical} and dynamic linear model frameworks that capture county-level spatio-temporal mortality patterns \citep{gibbs2020modeling}. These developments have enabled more sophisticated analyses of mortality patterns across geographic scales, from neighborhood-level studies \citep{wen2023modelling} to national-scale investigations \citep{dwyer2016us}.

\subsection{The Impact of Mental Health and Socio-economic Factors on Mortality}
\label{subsec:litreviewmental}

Suicide deaths have long been theorized to be caused by both internal factors (see \citealp{DeBeurs2019}) and external factors (see \citealp{Mueller2021}), with a complex interaction between individual psychological processes and broader sociocultural factors (see \citealp{Comtois2025}). For example, a recent review of metanalyses found that death by suicide was best predicted in part by involvement in the justice system, foster care experience, and unemployment (\citealp{Na2025}). Beyond individual level factors, county level characteristics have also been found to be predictive of suicide deaths. \cite{Liu2023} used an aggregate index of socioeconomic factors (including demographics, health care access, housing availability, education, unemployment rate, etc.) to predict county level suicide rates. They found that social vulnerability increased suicide rates between 56\% and 82\%. These studies illustrate two important considerations when predicting suicide deaths: first, predictions that incorporate internal and external processes are likely more powerful, and second, predictors can be found at both the individual and the community level.

Research also demonstrates a strong association between mental health disorders and increased mortality risk across multiple causes of death. A comprehensive systematic review and meta-analysis of 203 studies across 29 countries revealed that individuals with mental disorders face more than double the mortality risk of the general population, with mental disorders contributing to an estimated 14.3\% of deaths worldwide—approximately 8 million deaths annually \citep{walker2015mortality, ali2022excess}. This elevated mortality stems from both natural and unnatural causes, with cardiovascular diseases representing a major contributor to reduced life expectancy among individuals with severe mental illness and some finding that those with severe mental illness are twice as likely to die from infectious diseases   \citep{jayatilleke2017contributions, ronaldson2024severe}.

Notably, even suicidal ideation without suicide death appears to serve as an indicator of broader health vulnerability, as individuals experiencing such thoughts demonstrate increased mortality from natural causes \citep{batterham2013association, jonson2023life, shiner2016suicidal}. The COVID-19 pandemic has further highlighted these vulnerabilities, with patients having mental health disorders such as schizophrenia and bipolar disorder showing significantly higher COVID-19-related mortality compared to those without mental health conditions \citep{fond2021association, wang2020global}.

The geographic dimension of mental health and mortality has gained increasing attention in recent research. \citet{song2022spatially} employed Bayesian spatio-temporal models to examine associations between mental illness, substance use mortality, and unemployment across U.S. counties, revealing distinct regional risk clusters. Similar approaches have been applied internationally, with studies in South Korea \citep{kim2024spatially} using geographically weighted regression to identify spatial clusters of high suicide rates, and research in Chicago \citep{lotfata2023spatiotemporal} employing multilevel models to examine spatio-temporal associations between mental distress and socio-economic factors.

\citet{fontanella2018mapping} demonstrated the utility of Bayesian conditional autoregressive models for mapping suicide mortality in Ohio, finding strong associations between socio-economic deprivation, healthcare provider density, and suicide clusters. These spatial approaches have been enhanced through empirical Bayes methods that stabilize mortality estimates in sparsely populated areas \citep{manton1989empirical, clayton1987empirical}, addressing a key challenge in spatial mortality analyses.

Socio-economic factors are closely tied to both mortality and mental health patterns, with income, education, and employment status identified as having important relationships with mortality rates \citep{chetty2016association, barbieri2022socioeconomic}. Geographic variation in mortality has been extensively documented, with studies revealing substantial disparities across U.S. counties \citep{dwyer2016us} and highlighting how these disparities have evolved over time \citep{ezzati2008reversal}. The use of multilevel geographic approaches has proven essential, as analyzing health outcomes at only one geographic scale can obscure important patterns \citep{kim2016s, boing2020quantifying}.
International comparisons have shown that while mortality inequality exists globally, its patterns vary significantly by country and region \citep{mackenbach2008socioeconomic, atalay2023mortality}. These findings underscore the importance of considering local contextual factors when modeling mortality patterns and suggest that hierarchical Bayesian models are particularly well-suited for capturing these multi-level geographic effects.

From an actuarial perspective, the integration of mental health factors into mortality modeling represents both an opportunity and a challenge. Traditional actuarial models have increasingly incorporated socio-economic variables and epidemic effects \citep{biffis2005affine, cairns2024drivers}, but the systematic inclusion of mental health indicators remains limited. The COVID-19 pandemic has highlighted the need for more sophisticated mortality models that can account for the complex interactions between mental health, socio-economic factors, and infectious disease mortality \citep{delbrouck2024covid}.

Recent work has begun to address these challenges through the development of cause-specific mortality models that can be decomposed by contributing factors \citep{villegas2025modelling} and copula-based approaches that capture changing dependence structures across causes of death \citep{li2023covid}. These methodological advances provide a foundation for incorporating mental health variables into actuarial mortality models while maintaining computational tractability.

Despite significant progress in both spatial mortality modeling and mental health research, the integration of these fields remains underdeveloped. Most existing studies focus on either spatial / temporal mortality patterns or mental health associations with mortality, but few have attempted to create unified models that simultaneously capture spatial dependencies, temporal trends, and mental health effects. The development of such integrated models represents a significant opportunity to advance both academic understanding and practical applications in public health and actuarial science.

\section{Data}
\label{sec:data}

\subsection{Mortality Data}
\label{subsec:mortdata}

Mortality data for this study were obtained from the Division of Vital Statistics within the National Center for Health Statistics (NCHS), a division of the Centers for Disease Control and Prevention (CDC) \citep{data_nchs}. These data include demographic details and mortality records for all individuals who died in the United States between 2000 and 2023, disaggregated by year, age group, sex, and county of residence.

To estimate population-based mortality rates, population exposures for each demographic group were sourced from annual estimates provided by the U.S. Census Bureau. These estimates were derived from the most recent decennial census and adjusted annually to reflect changes due to births, deaths, and migration. Age was grouped into eighteen five-year intervals, ranging from 0–4 to 85 and older, and separate models were developed for each combination of age group and sex, resulting in a total of 36 models.

Several preprocessing steps were applied to prepare the data for modeling. Counties with population estimates of zero for a given age group, or (in rare cases) with more reported deaths than estimated population, were merged with the neighboring county with the largest population to preserve spatial structure while maintaining stable mortality estimates. (Throughout this study, the term ``county'' is used broadly to refer to sub-state geographic regions, including those in the state of Louisiana, which uses the alternative term ``parish'' for their sub-state subdivisions.)

To ensure consistency across the full 24-year span, county names and boundaries were standardized to reflect their configuration as of 2023. A small number of counties were adjusted accordingly, and a complete list of these adjustments is available in Appendix \ref{app:dataadjust}. Alaska, Hawaii, and U.S. territories were excluded from the analysis due to the difficulty in establishing meaningful spatial adjacency with the contiguous United States, resulting in 3100 counties before combining. After data cleaning and aggregation, the final dataset included 2974 counties. 

\subsection{Mental Health America Data}
\label{subsec:mentaldata}

While most individuals access Mental Health America (MHA) Screening organically, MHA has 200 affiliate organizations and multiple partner organizations that often refer users to the MHA Screening Program. Individuals who took a screening test and were referred by one of these affiliates were excluded from the dataset to reduce referral-related bias. In addition, counties with fewer than five positive screening results for a given screen and year were excluded to reduce instability in rate estimates. In this paper, a ``positive screening result'' is calculated based on the criteria defined for each instrument, as described below.

\begin{itemize}
  \item \textbf{Depression:} Data on severe depression were obtained from MHA’s Depression dashboard, which uses the Patient Health Questionnaire-9 (PHQ-9), a validated screening tool. The PHQ-9 includes nine items that assess the frequency of depressive symptoms over the past two weeks. Respondents rate each item on a four-point scale from ``not at all'' (0) to ``nearly every day'' (3). Severe depression is defined as a total score between 20 and 27, indicating frequent experience of multiple symptoms. The count of individuals meeting this threshold was reported by MHA.

  \item \textbf{Suicidal ideation:} These data came from MHA’s Suicide dashboard, which also uses the PHQ-9. Suicidal ideation is measured using item 9 of the PHQ-9: \textit{``Thoughts that you would be better off dead, or of hurting yourself.''} Individuals who responded ``more than half the days'' or ``nearly every day'' were classified as experiencing frequent suicidal ideation.

  \item \textbf{PTSD:} MHA’s PTSD dashboard reports results from the Primary Care PTSD Screen for DSM-5 (PC-PTSD-5), a five-item instrument. Respondents are asked whether they have experienced symptoms such as intrusive thoughts, avoidance, hypervigilance, emotional detachment, or guilt related to a traumatic event. A screen is considered positive if the respondent answers ``Yes'' to three or more of the five items.

  \item \textbf{Psychosis risk:} Data on clinical high risk for psychosis were obtained from MHA’s Psychosis dashboard, which is based on the Prodromal Questionnaire–Brief Version (PQ-B). This 21-item tool asks about unusual thoughts, feelings, or perceptions over the past month, with Yes/No response options. If ``Yes,'' respondents are prompted to rate the level of distress associated with the experience on a five-point Likert scale. These distress scores are summed, with scores of 24 or higher considered indicative of elevated risk for psychosis.
\end{itemize}

\begin{figure}[htbp]
  \centering
  \begin{subfigure}[b]{0.48\textwidth}
    \includegraphics[width=\linewidth]{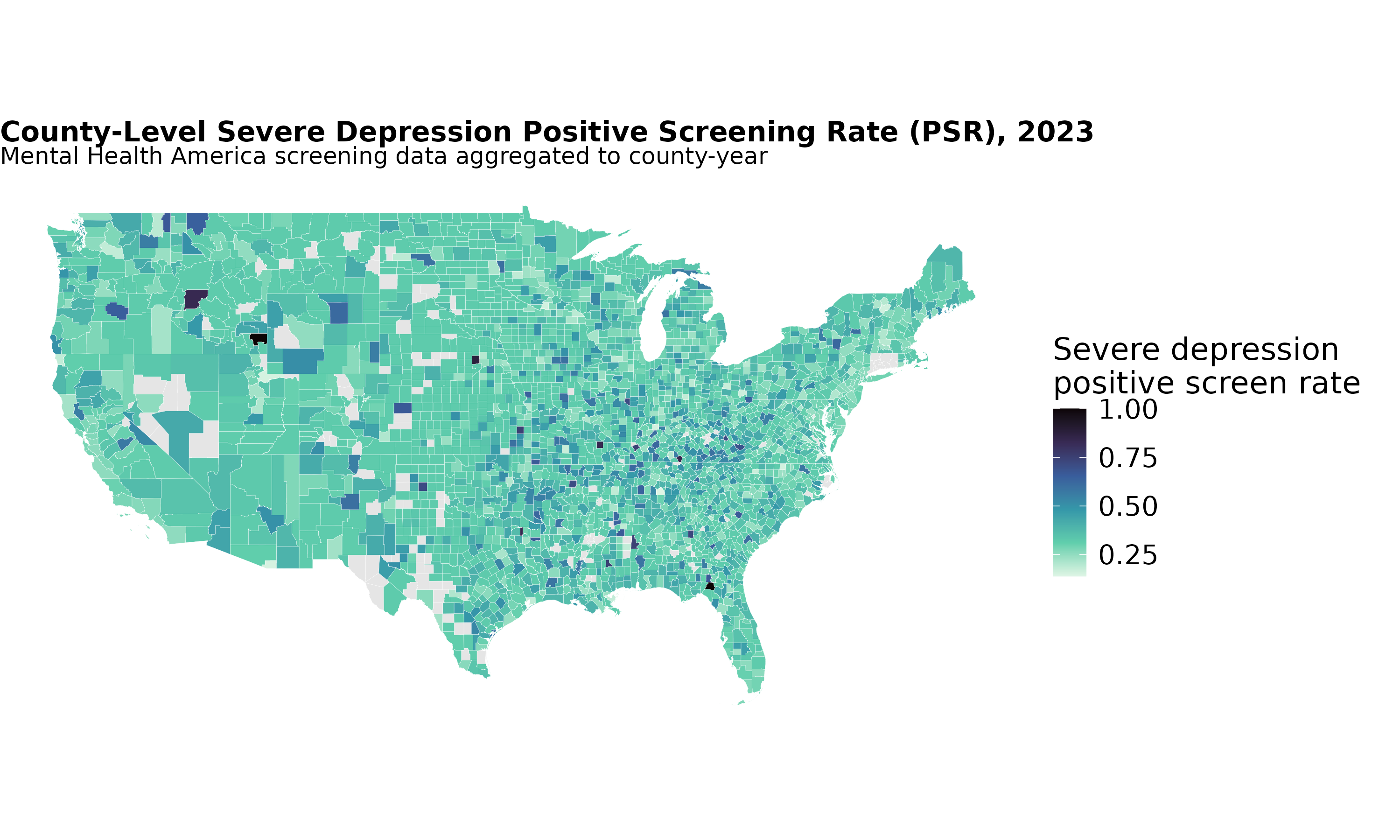}
    \caption{Severe depression PSR}
    \label{fig:psr_dep_map_2023}
  \end{subfigure}%
  \hfill
  \begin{subfigure}[b]{0.48\textwidth}
    \includegraphics[width=\linewidth]{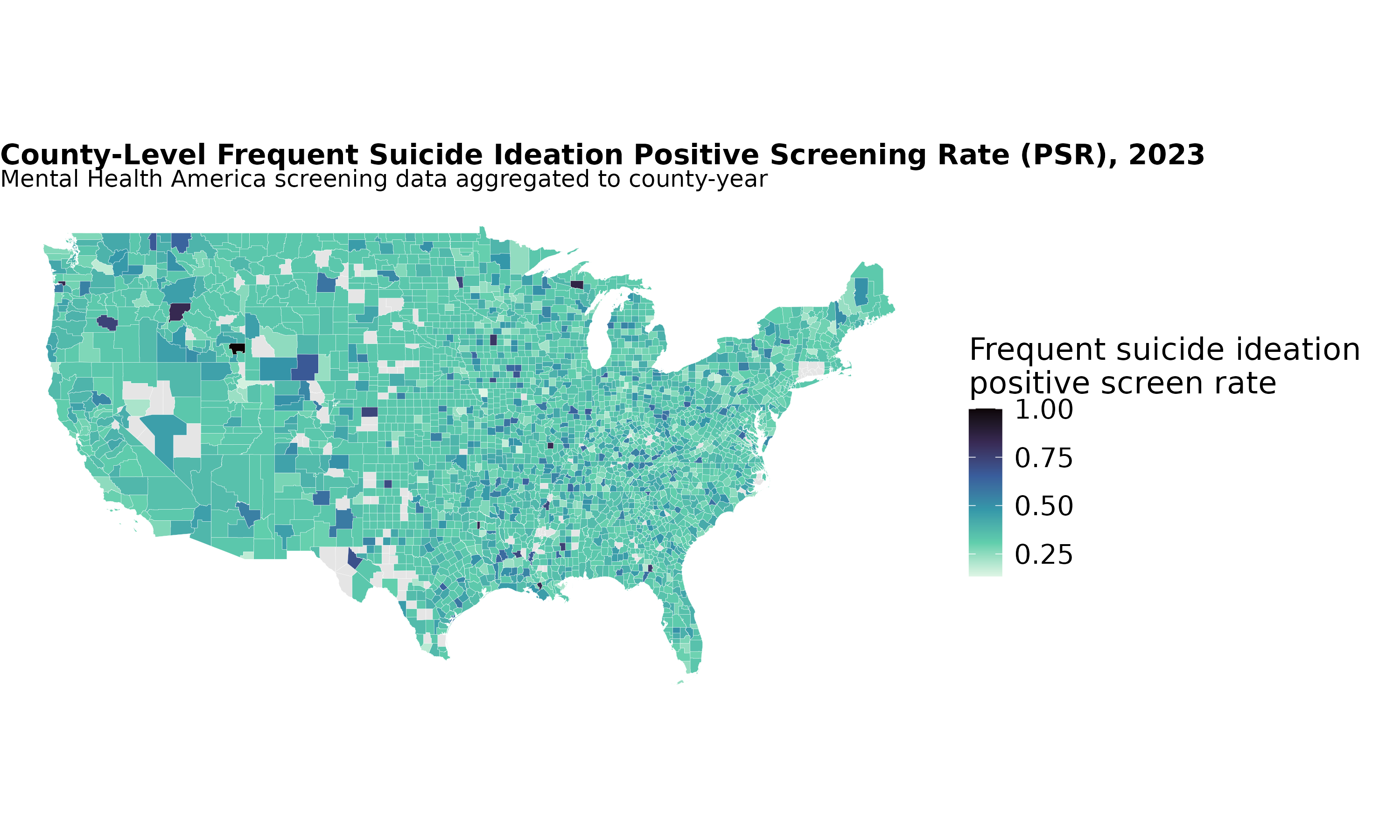}
    \caption{Frequent suicidal ideation PSR}
    \label{fig:psr_si_map_2023}
  \end{subfigure}

  \vspace{0.6em}

  \begin{subfigure}[b]{0.48\textwidth}
    \includegraphics[width=\linewidth]{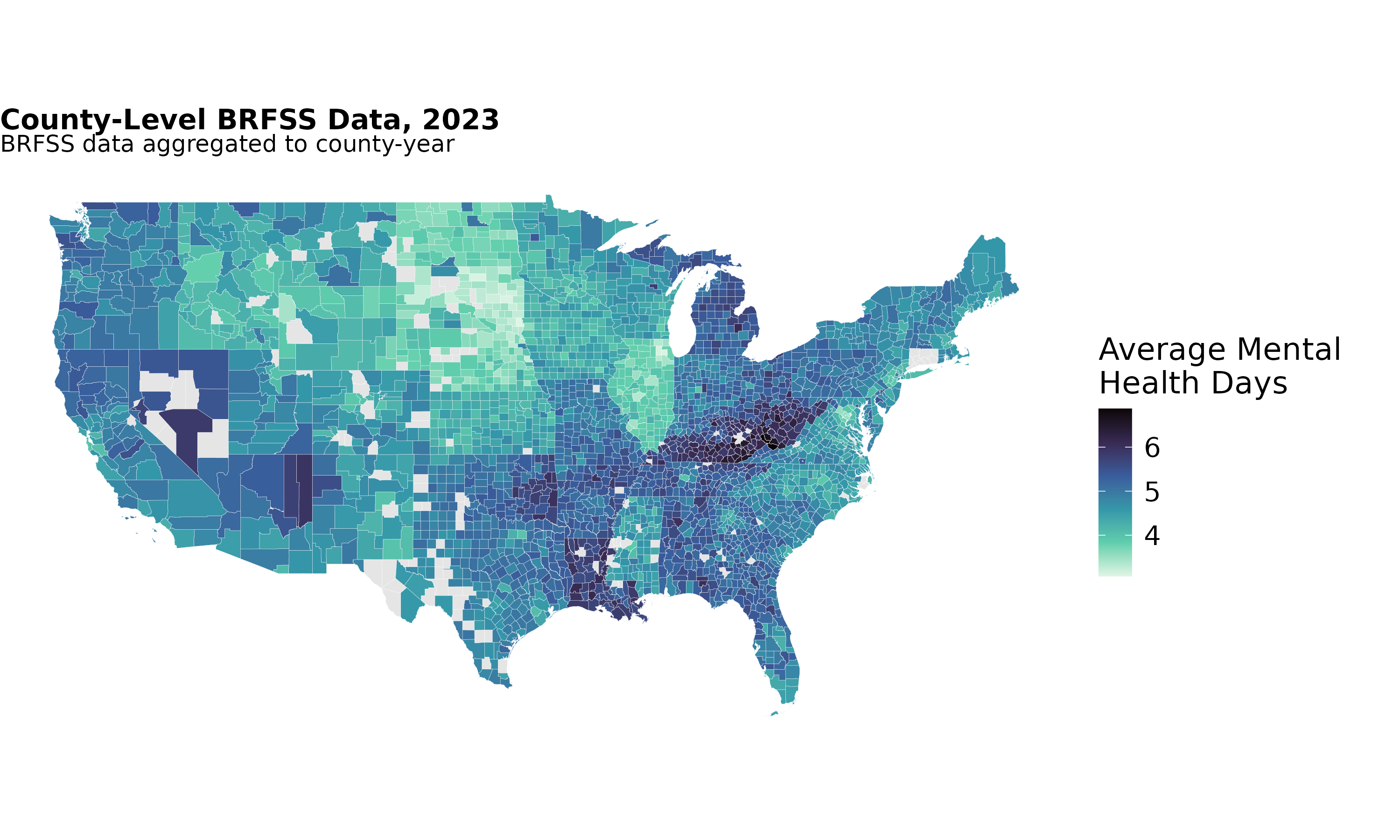}
    \caption{Frequent mental distress (BRFSS)}
    \label{fig:psr_brfss_map_2023}
  \end{subfigure}%
  \hfill
  \begin{subfigure}[b]{0.48\textwidth}
    \centering
    \mbox{}
  \end{subfigure}

  \caption{County-level mental-health indicators in 2023:  
  (a) severe depression, (b) frequent suicidal ideation, and (c) frequent mental distress.  
  Counties with insufficient data are shaded light grey.}
  \label{fig:mental_maps_4panel}
\end{figure}

To construct our model covariates, we compute the Positive Screening Rate (PSR) for each screen, defined as the number of individuals who screened positive divided by the total number of completed screens for that instrument, by county and year. For example, if 10 participants in Maricopa County screened positive for severe depression in 2020 out of 1{,}000 total depression screens, the PSR would be 0.01. We calculate analogous rates for each of the mental health outcomes described above and use these values as county-level covariates in our model. Figures \ref{fig:psr_dep_map_2023} and \ref{fig:psr_si_map_2023} compare county-level PSRs for severe depression and frequent suicidal ideation in 2023.

\subsection{Behavioral Risk Factor Surveillance System (BRFSS) and PLACES Data}

In addition to screening data from Mental Health America, we incorporated county-level estimates of self-reported mental health from the Behavioral Risk Factor Surveillance System (BRFSS), as distributed through the \textit{PLACES} project \citep{cdc_places}. BRFSS is a nationally representative, state-based telephone survey conducted by the Centers for Disease Control and Prevention (CDC) that collects detailed information on health-related risk behaviors, chronic health conditions, and use of preventive services \citep{brfss_design}.

We specifically used the PLACES dataset, which provides model-based estimates of population health indicators for U.S. counties and census tracts. These estimates are generated through a multi-level regression and poststratification approach that combines BRFSS survey data with demographic and geographic information from the U.S. Census and American Community Survey (ACS). This modeling approach allows for more reliable small-area estimation, particularly in sparsely sampled regions \citep{places_methodology}.

From this dataset, we extracted the indicator corresponding to the proportion of adults in each county who reported having 14 or more days of poor mental health in the past 30 days, a variable referred to as \texttt{MENTHLTH\_GE14D}. This variable is derived from a count question in the original BRFSS survey, which asks respondents to report the number of days (0–30) during which their mental health was not good. PLACES dichotomizes this count at a clinically meaningful threshold of 14 days to indicate \textit{frequent mental distress}, a measure widely used in public health surveillance and epidemiologic studies \citep{mmwr_fmd}.

PLACES data used in this study were drawn from the 2023 release and are based on BRFSS responses collected in 2021. These model-based estimates complement our screening-derived indicators by capturing a broader population-representative perspective on mental health burden. Figure \ref{fig:psr_brfss_map_2023} displays model-based 2023 estimates of the proportion of respondents whose number of poor-mental-health days exceeded the BRFSS threshold.

\subsection{Socio-economic Data}
\label{subsec:covdata}

We considered several socio-economic variables in our analysis, as many of them have previously been shown to be related to mortality and mental health, as discussed in Section \ref{subsec:litreviewmental}: 
    \begin{itemize}
        \item \textbf{Alcohol Consumption:} Per capita alcohol consumption data was collected at the state level for each year by the National Institute on Alcohol Abuse and Alcoholism at the National Institutes of Health \citep{slater2024surveillance}.
        \item \textbf{Education:} Level of educational attainment was collected at the county level for each year by the U.S. Census Bureau as a part of the American Community Survey \citep{census2023acs}. We used the percentage of adults with at least a bachelor's degree as our measure of educational level.
        \item \textbf{House Price Index:} House price indices (HPI) were calculated at the state level for each year by the Federal Housing Finance Agency, based mortgage data from Fannie Mae and Freddie Mac \citep{fhfa2024hpi}.
        \item \textbf{Marital Status:} The percentage of adults who were married was collected at the county level for each year by the U.S. Census Bureau as a part of the American Community Survey \citep{census2023acs}. 
        \item \textbf{Household Size:} The average household size was collected at the county level for each year by the U.S. Census Bureau as a part of the American Community Survey \citep{census2023acs}.
        \item \textbf{Unemployment:} Unemployment rate was collected at the county level for each year by the U.S. Bureau of Labor Statistics \citep{bls2023laus}.
        \item \textbf{Race:} To measure the impact of race, we used the percentage of heads of households who identify as white. This data was collected at the county level for each year by the U.S. Census Bureau as a part of the American Community Survey \citep{census2023acs}.
    \end{itemize}

To address missing and inconsistent data prior to analysis, additional preprocessing steps were applied. For the socio-economic data, we used linear interpolation to fill in missing years. In the MHA survey data, counties with missing values for either the number of positive responses or total responses were imputed using the national average, calculated from all available, non-missing county-level responses. Counties with insufficient mortality data after this process are depicted in gray in the exploratory data analysis (EDA) maps, indicating the absence of reliable estimates.

\subsection{Exploratory Data Analysis of Suicide Trends}

\begin{figure}[htbp]
    \centering
    \includegraphics[width=0.8\textwidth]{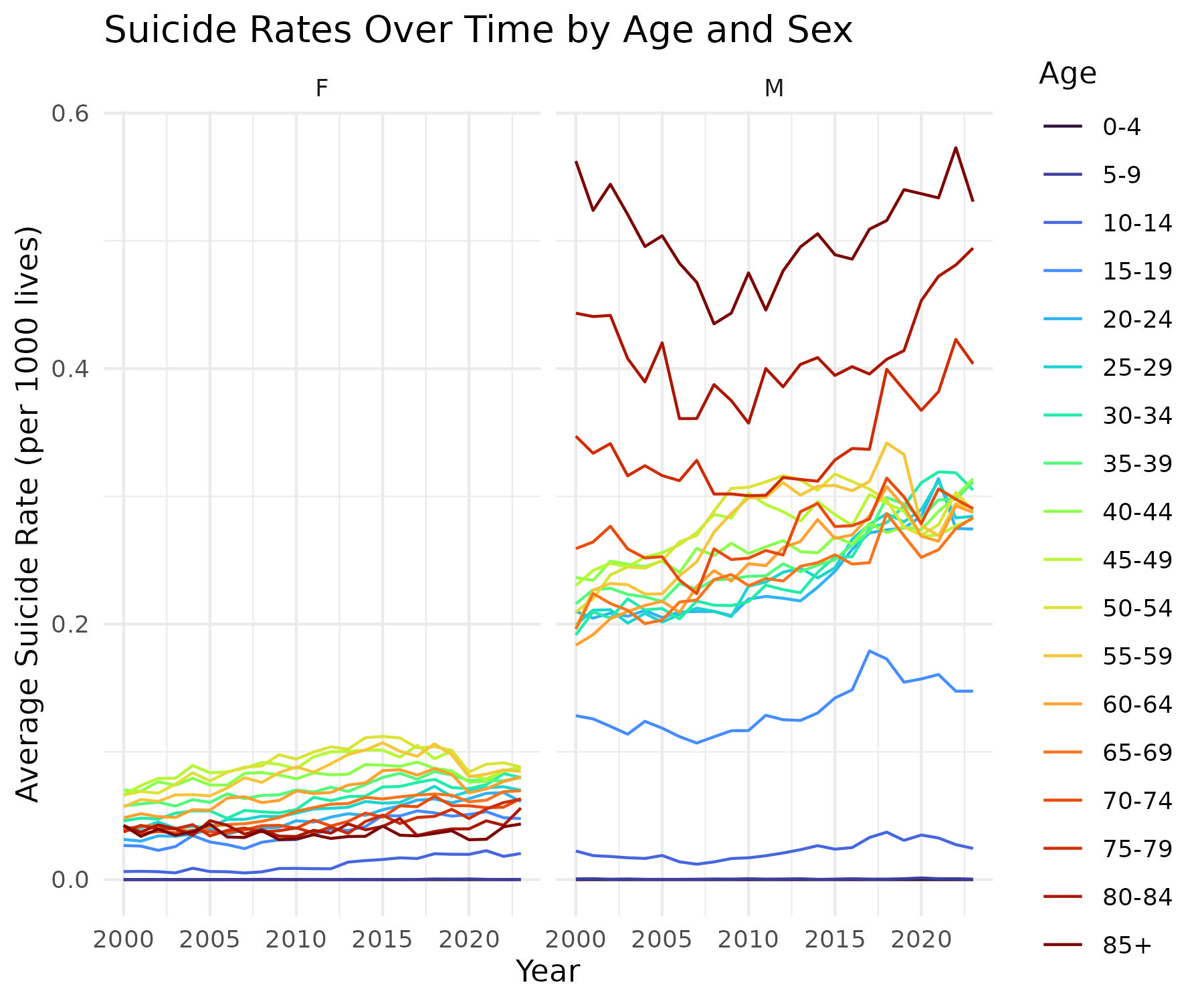}
    \caption{Average suicide rates over time, stratified by sex and five-year age group. Trends in age- and sex-specific suicide rates in the United States, 2000--2023, based on mortality data from the National Center for Health Statistics. Rates are expressed as deaths per person-year within each age-sex group.}
    \label{fig:suicide_rates_combined}
\end{figure}

To contextualize the subsequent modeling, we examined temporal trends in age- and sex-specific suicide rates from 2000 to 2023. Figure~\ref{fig:suicide_rates_combined} presents average suicide rates stratified by sex and five-year age groups. (Figures \ref{fig:suicide_rates_45_down} and \ref{fig:suicide_rates_45_up} in Appendix \ref{app:suppplots} give the same information split by age range for additional clarity.)

For females, the highest suicide rates occur between the ages of 40–59. The rates are very close to zero for girls under 10 years old, with an increase in rates during the teenage years. For many of the age groups, there appears to be an increasing trend over the observed time period. Many age groups experienced a decline in suicide rates in the year 2020, the same year as the inception of the COVID-19 pandemic.

The highest suicide rates are among older males (75+). For those ages, the suicide rates decreased between 2000 and 2009 and have been increasing since then. For all ages older than 15, female suicide rates are much smaller than male suicide rates. Male rates are essentially zero for children and then increase for teenagers and increase further for young adults. The rates are pretty similar for adult males (20-75), with a pretty consistent increasing pattern throughout our study period.

\section{Models and Methods}
\label{sec:methods}

To model spatio-temporal trends in mental health outcomes across counties in the continental United States, we use a Bayesian hierarchical model structure similar to that utilized by \citet{gibbs2020modeling}. The model accounts for both spatial correlation between neighboring counties and linear temporal trends within each county. The model imposes a first-order Markov property on the spatial structure, whereby each region is conditionally independent of all non-adjacent regions given its neighboring counties. We use this same underlying statistical model for both parts of our analysis (described in more detail in Section \ref{sec:twoModels}).

\subsection{Model Details}

Let $\bf{Y}$ $= \{y_{kt}\}$ be a vector of data points that we are modeling, i.e., either all-cause deaths or deaths by suicide. For simplicity and ease of exposition, we describe the modeling of $\bf{Y}$ in terms of deaths, but the same model applies to our analysis of suicides. Specifically, let $y_{kt}$ represent the number of deaths in county $k$, where $k=1,2,\dots,K$ and time $t$, where $t=1,2,\dots,T$. Since the data are discrete counts and we assume the populations to be known, we can assume a binomial likelihood with
\(y_{kt} \sim \text{Binomial}(m_{kt}, \theta_{kt})\) 
where $m_{kt}$ is the known population and $\theta_{kt}$ is the unknown probability of death in county $k$ at time $t$. 

We utilize a logistic link function to achieve this likelihood on the probability of death:
\(\log(\theta_{kt}/(1-\theta_{kt}))=\bf{x_{kt}^\top}\boldsymbol{\beta}+\psi_{kt},\)
where $\bf{x_{kt}}$ is a vector of observed covariates, including socio-economic data or measures of mental health burden (depression, suicide ideation, PTSD, and psychosis) derived from population screening tools, standardized by the total number of screenings., i.e., the PSRs defined in Section \ref{subsec:mentaldata}. The vector $\boldsymbol{\beta}$ contains the fixed-effect coefficients representing the effect of each covariate on mortality risk. The term $\psi_{kt}$ represents a spatio-temporal random effect used to account for unmeasured spatial and temporal variation in mortality risk for county $k$ at time $t$.

This random effect $\psi_{kt}$ is decomposed into spatial and temporal components, 
\(\psi_{kt}=\phi_k+(\alpha+\delta_k)\cdot \frac{t-\bar{t}}{T},\)
where $\phi_k$ is a spatial random effect representing the baseline mortality level in county $k$, $\alpha$ is a national (overall) time trend, and $\delta_k$ is a county-specific deviation from that trend. This latter term allows for an interaction between spatial and temporal effects. Both $\boldsymbol{\phi} = \{\phi_1, \dots,\phi_k\}^\top$ and $\boldsymbol{\delta} = \{\delta_1, \dots,\delta_k\}^\top$ are assigned conditional autoregressive (CAR) priors to encourage spatial smoothing, meaning counties that are geographically adjacent are modeled to have similar baseline mortality rates and trends over time.

The CAR prior is parameterized such that
\(\boldsymbol{\phi} \sim \mathcal{N}_K(0, \tau_\phi^2(\mathbf{D}-\rho_{\text{int}} \mathbf{W})^{-1}),\)
where $\mathbf{W}$ is the symmetric $K \times K$ adjacency matrix of ones and zeros indicating neighboring counties, $\mathbf{D}$ is a $K \times K$ diagonal matrix with each entry along the main diagonal equal to the number of neighbors of the given county, $\tau_\phi^2$ is the spatial variance parameter, and $\rho_{\text{int}}$ controls the strength of spatial dependence. A similar prior is placed on $\boldsymbol{\delta}$; that is,
\(\boldsymbol{\delta} \sim \mathcal{N}_K(0, \tau_\delta^2(\mathbf{D}-\rho_{\text{slo}} \mathbf{W})^{-1}),\) where the $\tau_\delta^2$ and $\rho_{\text{slo}}$ parameters are the temporal analogs to $\tau_\phi^2$ and $\rho_{\text{int}}$, respectively.

Estimation was performed in a fully Bayesian framework. We specified weakly informative, independent priors for the model parameters. The prior distribution for the fixed effects $\boldsymbol{\beta}$ was a multivariate normal with mean vector $\boldsymbol{\mu}_\beta=0$ and covariance matrix $\mathbf{\Sigma}_\beta=\textit{I}$. The prior for the national time trend parameter $\alpha$ was also normal, with mean $\mu_\alpha=0$ and variance $\sigma^2=1000^2$, reflecting minimal prior information. For the variance components associated with the spatial and temporal random effects ($\tau_\delta^2$ and $\tau_\phi^2$), we used inverse-gamma (IG) priors with shape 1 and rate 0.01.
The correlation parameters $\rho_{\text{int}}$ and $\rho_{\text{slo}}$ were assigned uniform priors on $[0, 1].$ Thus, the full Bayesian model is given by
\begin{align*}
&y_{kt} \sim \text{Binomial}(m_{kt}, \theta_{kt}) \\
\text{logit}(\theta_{kt}) &= \mathbf{x}_{kt}^\top \boldsymbol{\beta} + \psi_{kt}, ~~~~
\psi_{kt} = \phi_k + (\alpha + \delta_k) \cdot \frac{t - \bar{t}}{T} \\
\boldsymbol{\phi} &\sim \text{CAR}(\rho_{\text{int}}, \tau_\phi^2), ~~~
\boldsymbol{\delta} \sim \text{CAR}(\rho_{\text{slo}}, \tau_\delta^2) \\
\alpha &\sim \mathcal{N}(0, 1000^2),
 ~~~ \boldsymbol{\beta} \sim \mathcal{N}(\bf{0}, 1000^2 \cdot \bf{I}) \\
\tau_\phi^2, \tau_\delta^2 &\overset{iid}{\sim} \text{IG}(1, 0.01), 
~~~\rho_{\text{int}}, \rho_{\text{slo}} \overset{iid}{\sim} \text{Uniform}(0, 1)
\end{align*}

Under this choice of model and prior distributions, only the spatial and temporal random effect parameters ($\tau_\delta^2$ and $\tau_\phi^2$) have closed forms for their full conditional distributions, given the other parameter values:

\[ \tau_\phi^2 \sim \text{IG}(1.5, 0.01 + \boldsymbol{\phi}^\top(D- \rho_{\text{int}} W)\boldsymbol{\phi}/2)\]
\[ \tau_\delta^2 \sim \text{IG}(1.5, 0.01 + \boldsymbol{\phi}^\top(D- \rho_{\text{slo}} W)\boldsymbol{\phi}/2)\]

For the other parameters, posterior inference was conducted using a Metropolis-within-Gibbs algorithm. In each iteration, we alternately sampled the fixed effects vector $\boldsymbol{\beta}$, the national time trend parameter $\alpha$, and the spatio-temporal random effects $\boldsymbol{\phi}$ and $\boldsymbol{\delta}$ using Metropolis-Hastings (MH) updates with Gaussian random walk proposals. For each parameter block, proposal variances were tuned adaptively during the burn-in phase to achieve target acceptance rates between 0.3 and 0.5. The spatial and temporal correlation parameters $\rho_{\text{int}}$ and $\rho_{\text{slo}}$ were also updated via MH steps, with proposal values constrained to the interval $(0,1)$ using logistic transforms. For all of our models, we utilized the \texttt{CARBayesST} package \citep{lee2018spatio} in \textsf{R} \citep{baseR}. See Table \ref{tab:model-structure} for a summary of the model parameters.

\begin{table}[!ht]
\centering
\caption{Model structure parameters and their roles (all age $\times$ sex strata).}
\label{tab:model-structure}
\renewcommand{\arraystretch}{1.1}
\begin{tabularx}{\textwidth}{p{4cm} p{2cm} X X}
\hline
\textbf{Quantity} & \textbf{Symbol} & \textbf{Meaning / Interpretation} \\
\hline
Intercept & $\beta_{0}$ & Baseline log-odds of death in a county-year, after covariate adjustment. \\
National time trend & $\alpha$ & Average change in log-odds per year (centered time). \\
County time deviation & $\delta_k$ & County-specific deviation from the national time trend. \\
Spatial random effect & $\phi_k$ & County baseline deviation from the national level after covariates. \\
Spatio-temporal RE & $\psi_{kt}=\phi_k+(\alpha+\delta_k)\frac{t-\bar t}{T}$ & Combined spatial baseline and (national $+$ county) linear time component. \\
Spatial dependence & $\rho_{\text{int}}$ & Strength of spatial clustering in $\phi$ (CAR prior). \\
Temporal dependence & $\rho_{\text{slo}}$ & Similarity of county time slopes across neighbors (CAR prior on $\delta$). \\
Spatial variance & $\tau^2_{\phi}$ & Marginal variance of $\phi$ under the CAR prior. \\
Temporal variance & $\tau^2_{\delta}$ & Marginal variance of $\delta$ under the CAR prior. \\
Outcome (counts) & $y_{kt}$ & Binomial likelihood with exposure $m_{kt}$ and $\theta_{kt}$; logit link. \\
Probability of death & $\theta_{kt}$ & Modeled via $\text{logit}(\theta_{kt})=x^\top_{kt}\beta+\psi_{kt}$. \\
\hline
\end{tabularx}
\end{table}

\subsection{Modeling Approach for Research Questions}
\label{sec:twoModels}

In this paper, we address two broad questions:  
\begin{enumerate}
    \item What are the impacts of socio-economic variables and spatial relationships on suicides?
    \item What is the impact of county-level mental health on county-level total deaths and suicides (after accounting for spatial correlation)?
\end{enumerate}

To answer these questions, we fit two distinct sets of spatio-temporal models.  
For Question 1, we leverage socio-economic and behavioral covariates over a long historical period to examine their relationship with suicide mortality.  
For Question 2, we explore multiple measures of mental health, from both short-term direct screening data and long-term survey-based estimates, to assess their association with both all-cause mortality and suicide mortality.  

Both sets of models use the hierarchical spatio-temporal framework described in the Methods section, with binomial likelihoods for mortality counts and county-year–indexed spatial and temporal random effects. The primary differences lie in the outcome variable, covariate set, and temporal coverage of the data.

\subsubsection{Socio-economic Models of Suicide}

To investigate Question 1, we estimate suicide mortality rates using annual counts of suicides in each county. Our motivation for this model is to assess the influence of socio-economic factors on suicide outcomes across a long study period.  

We selected covariates from the publicly available sources listed in Section~\ref{subsec:covdata}, covering 2010–2023, and supplemented these with the \textit{Mental Health Days} variable from the Behavioral Risk Factor Surveillance System (BRFSS). The BRFSS is a government survey conducted annually since 2010, which allowed us to incorporate over a decade of data. 

To capture the potential impact of major societal shifts, we included indicator variables for the COVID-19 pandemic period (2020–2021) and the post-pandemic period (2022–2023). This enables us to examine whether relationships between socio-economic factors and suicide mortality shifted during or after the pandemic.

The fixed-effects portion of the model is:
\[
\begin{aligned}
\mathbf{x}^\top\boldsymbol{\beta} = &\ \beta_0 
+ \beta_1\textit{Alcohol}_{kt}
+ \beta_2\textit{Educ}_{kt}
+ \beta_3\textit{Crime}_{kt}
+ \beta_4\textit{HPI}_{kt} \\
&+ \beta_5\textit{Married}_{kt}
+ \beta_6\textit{HHSize}_{kt}
+ \beta_7\textit{Unemp}_{kt}
+ \beta_8\textit{Race}_{kt} \\
&+ \beta_9\textit{MH\_Days}_{kt}
+ \beta_{10}\textit{COVID}_{t}
+ \beta_{11}\textit{Post 2021}_{t},
\end{aligned}
\]
where $k$ indexes counties and $t$ indexes years. Spatial and temporal random effects capture residual correlation not explained by the covariates.

\subsubsection{Mental Health Models of Total Deaths and Suicide}

To address Question 2, we fit two related models: one with all-cause mortality as the outcome, and another with suicide mortality as the outcome. This dual-outcome approach allows us to compare the role of mental health indicators in explaining broad mortality trends versus suicide-specific patterns.

Covariates were drawn from Mental Health America (MHA) data as described in the Data section, supplemented with selected behavioral health measures from the Behavioral Risk Factor Surveillance System (BRFSS).
By combining these sources, we can assess whether associations between mental health indicators and mortality outcomes are consistent across short-term direct measures (MHA) and longer-term survey-based estimates (BRFSS). This also enables sensitivity analyses to determine whether relationships are robust to differences in data source, coverage, and measurement methodology.

Our primary covariate of interest is the positive screen rate ($\textit{PSR}_{kt}$) for a given mental health condition in county $k$ and year $t$. The fixed-effects portion of the model is:
\[
\mathbf{x}^\top\boldsymbol{\beta} = \beta_0 + \beta_1\textit{PSR}_{kt},
\]
with spatial and temporal random effects accounting for residual correlation across counties and over time. See Table \ref{tab:covariate-map} for a summary of the covariates and associated coefficients across both model types.

\subsubsection{Computational Considerations}



Both models were run with a burn-in of 100,000 iterations, followed by 50,000 saved posterior samples.  To assess convergence and mixing, we visually inspected trace plots for all parameters, ensuring adequate exploration of the posterior distribution. We also computed the effective sample size (ESS) for each parameter and verified that all key parameters exceeded an ESS threshold of 200, indicating low autocorrelation and stable estimates. Finally, we used the Geweke diagnostic statistic to assess convergence for each chain; all parameters passed at a 5\% significance level, suggesting that the posterior means of early and late segments of the chains were consistent.

\section{Results and Discussion}
\label{sec:results}

Here we present the results of our analyses and discuss some of the most important implications of these findings. These results allow us to assess the level of spatial and temporal dependence among counties and to quantify the impact on all-cause and suicide-specific mortality of the various mental health and socio-economic variables described above. We consider these effects for males and females separately, and across all ages groups; this is important, since the effects often vary strongly by both age and sex. 

\subsection{Socio-economic Effects on Suicide}
\label{socecosui}

In this section, we present the estimated parameters from our spatio-temporal model using suicide as the target variable and several socio-economic indicators as covariates. We start by examining model structure parameters and explore what they indicate regarding the nature of the data by age group, time, and location. Then we examine more closely the specific covariate effects. In all, 36 models were fit, one for every age group and gender combination.

\subsubsection{Temporal and Spatial Trends}
\label{tempspat}

\begin{figure}[htbp]
  \centering
  \begin{subfigure}[t]{0.48\textwidth}
    \centering
    \includegraphics[width=\textwidth]{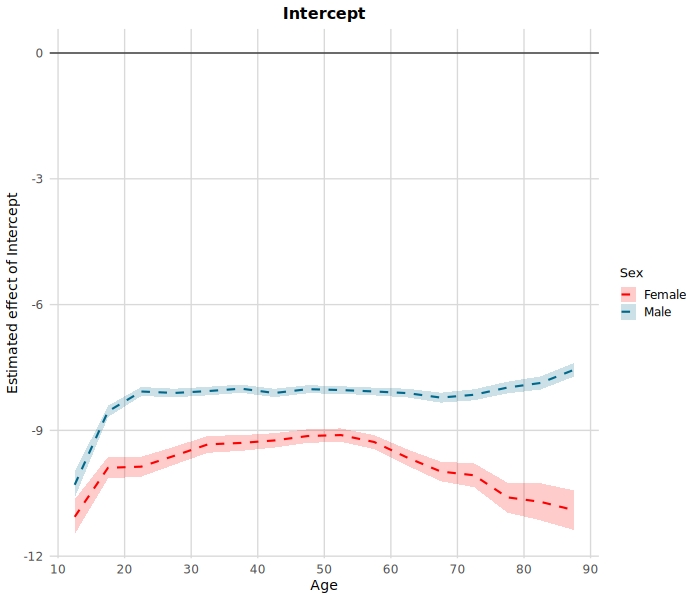}
    \caption{Estimates and 95\% credible intervals for (\(\beta_0\)) (intercept) by sex and age.}
    \label{fig:se_Intercept}
  \end{subfigure}%
  \hfill
  \begin{subfigure}[t]{0.48\textwidth}
    \centering
    \includegraphics[width=\textwidth]{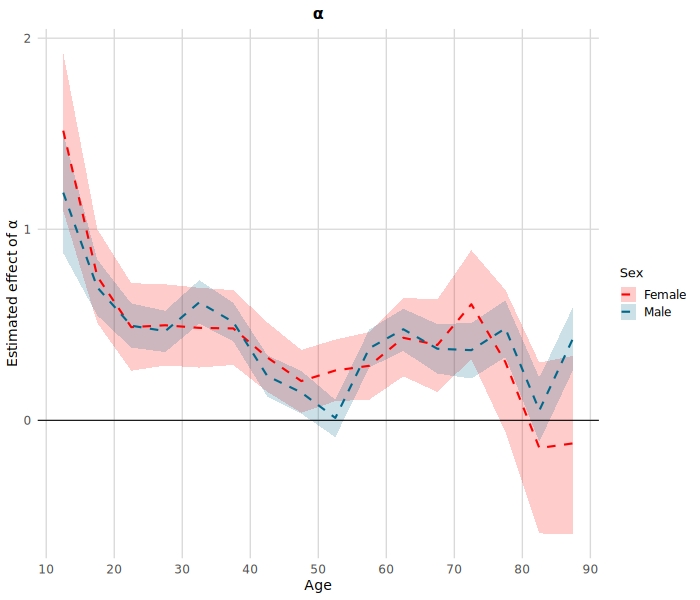}
    \caption{Estimates and 95\% credible intervals for (\(\alpha\)) (country-wide time trend) by age and sex.}
    \label{fig:se_alpha}
  \end{subfigure}
  
  \caption{Comparison of intercept (\(\beta_0\)) and time trend (\(\alpha\)) estimates by age and sex.}
  \label{fig:side_by_side}
\end{figure}

The intercept ($\beta_0$) comparison plot, Figure \ref{fig:se_Intercept}, illustrates the general pattern of suicide rates by age and sex. In all age groups, men consistently exhibit higher suicide rates than women. Among men, rates remain relatively high throughout adulthood and show a slight increase after age 70. Among women, suicide rates gradually increase until around age 50 and then gradually decline with increasing age.

 The model parameter $\alpha$, representing the national time trend, reveals whether suicide rates have increased or decreased over time. Positive values indicate increasing rates, while negative values indicate decreases. In Figure \ref{fig:se_alpha} we observe that in general, the results suggest that suicide rates have worsened over time, with the highest increases observed among younger age groups. For women 80 years and older, there is some indication of improvement, consistent with the patterns seen in the $\alpha + \delta_k$ maps. For men, dips in the values of $\alpha$ are visible in the age groups 45-54 and 80-84.

\begin{figure}[htbp]
  \centering
  \begin{subfigure}[t]{0.32\textwidth}
    \centering
    \includegraphics[width=\textwidth]{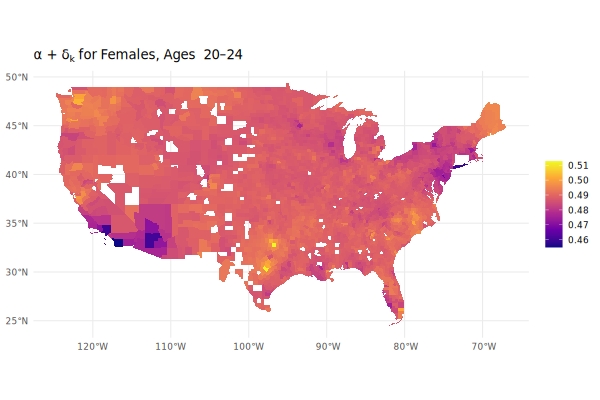}
    \caption{Females, 20--24}
    \label{fig:apd_f_20_24}
  \end{subfigure}%
  \hfill
  \begin{subfigure}[t]{0.32\textwidth}
    \centering
    \includegraphics[width=\textwidth]{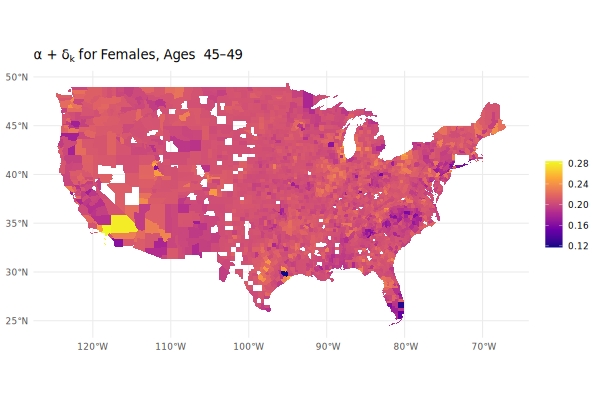}
    \caption{Females, 45--49}
    \label{fig:apd_f_45_49}
  \end{subfigure}%
  \hfill
  \begin{subfigure}[t]{0.32\textwidth}
    \centering
    \includegraphics[width=\textwidth]{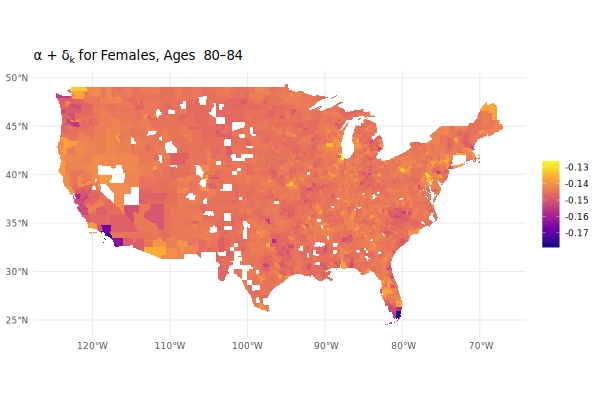}
    \caption{Females, 80--84}
    \label{fig:apd_f_80_84}
  \end{subfigure}

  \vspace{0.6em}

  \begin{subfigure}[t]{0.32\textwidth}
    \centering
    \includegraphics[width=\textwidth]{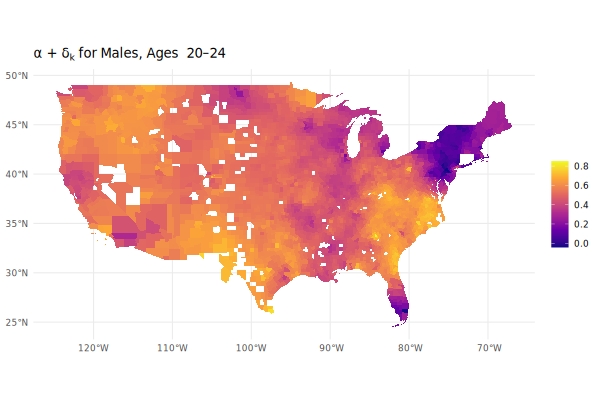}
    \caption{Males, 20--24}
    \label{fig:apd_m_20_24}
  \end{subfigure}%
  \hfill
  \begin{subfigure}[t]{0.32\textwidth}
    \centering
    \includegraphics[width=\textwidth]{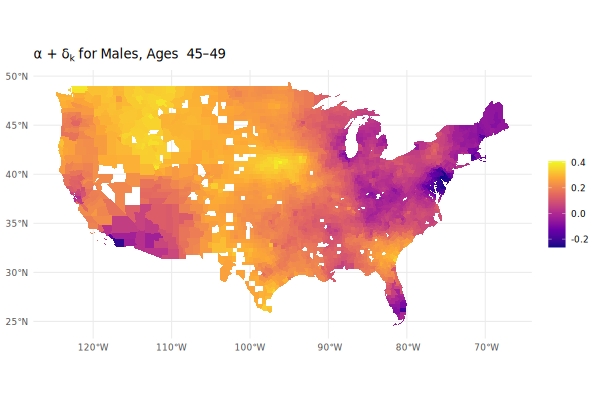}
    \caption{Males, 45--49}
    \label{fig:apd_m_45_49}
  \end{subfigure}%
  \hfill
  \begin{subfigure}[t]{0.32\textwidth}
    \centering
    \includegraphics[width=\textwidth]{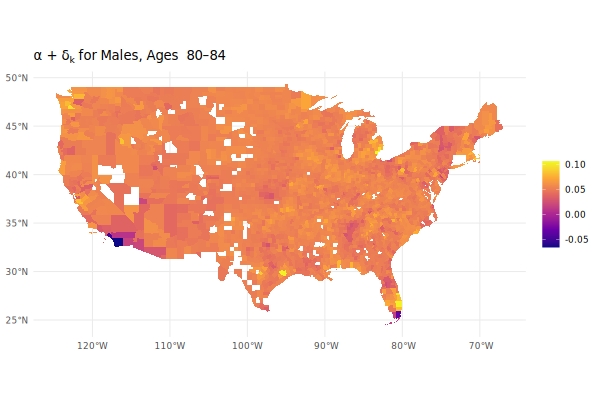}
    \caption{Males, 80--84}
    \label{fig:apd_m_80_84}
  \end{subfigure}

  \caption{County-specific time trends (\(\alpha + \delta_k\)) for ages 20--24, 45--49, and 80--84 by sex. Panels show deviations from the overall national trend, where higher values indicate counties with greater model-estimated suicide risk relative to the national average after accounting for covariates.}
  \label{fig:alphaplusdelta_6panel}
\end{figure}

The county-specific trends, $\alpha + \delta_k$, highlight where the local suicide trends diverge from the national pattern. Figure~\ref{fig:alphaplusdelta_6panel} shows these patterns for several age groups. Figure~\ref{fig:alphaplusdelta_6panel}\subref{fig:apd_f_80_84} shows the decline in suicide rates for those aged 80–84 which we see in Figure \ref{fig:se_alpha}, with all other age groups showing increased rates since 2010. For men, we see some significant regional variation, especially at the younger and middle ages, as is seen in Figure~\ref{fig:alphaplusdelta_6panel}\subref{fig:apd_m_20_24} and \ref{fig:alphaplusdelta_6panel}\subref{fig:apd_m_45_49}. There is a mixture of improving and worsening counties with typically the northeast, southern California, and southern Florida experiencing consistently lower rates of change than other parts of the country. For females and for older males, the rate of change is much more consistent across age groups. Some of the highest increases, particularly for men aged 45–49, are found in the West and in the region often referred to as the suicide belt \citep{wray2012suicide, harper2008there, RuralHealth2025, SPRC2024}.

Spatial correlation values ($\rho_\text{int}$), which quantify the degree to which neighboring counties have similar overall mortality levels, are close to 1 for essentially all age–sex groups with only a very slight drop for the 20–24 age group. (See Figure \ref{fig:se_rho_int} in Appendix \ref{app:suppplots}.) This demonstrates that there is a high magnitude of dependence in overall suicide rates between neighboring counties.

The temporal correlation ($\rho_{slo}$), which measures how similar neighboring counties are in their time trends, generally centers around 0.5 as seen in Figure \ref{fig:se_rho_slo} in Appendix \ref{app:suppplots}. Notable deviations include high correlations approaching 1 for males aged 15–29 and 35–44, and low correlations below 0.35 for females aged 60-65 and 80-85, as well as males aged 80-85. 

\begin{figure}[htbp]
  \centering
  \begin{subfigure}[t]{0.32\textwidth}
    \centering
    \includegraphics[width=\textwidth]{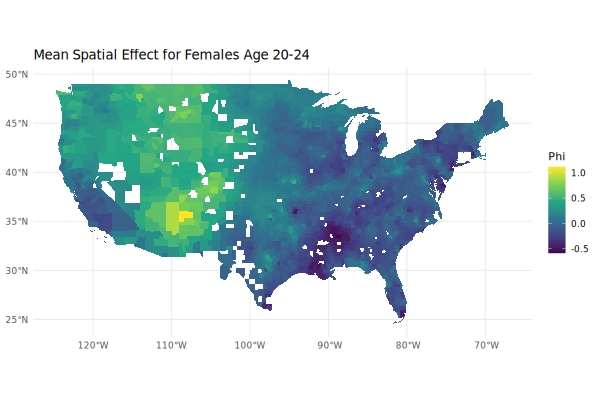}
    \caption{Females, 20--24}
    \label{fig:phi_f_20_24}
  \end{subfigure}%
  \hfill
  \begin{subfigure}[t]{0.32\textwidth}
    \centering
    \includegraphics[width=\textwidth]{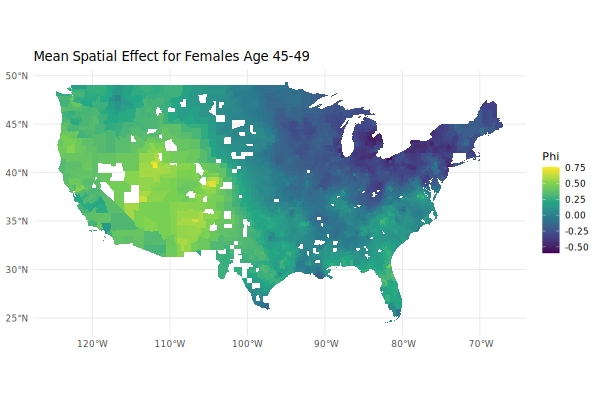}
    \caption{Females, 45--49}
    \label{fig:phi_f_45_49}
  \end{subfigure}%
  \hfill
  \begin{subfigure}[t]{0.32\textwidth}
    \centering
    \includegraphics[width=\textwidth]{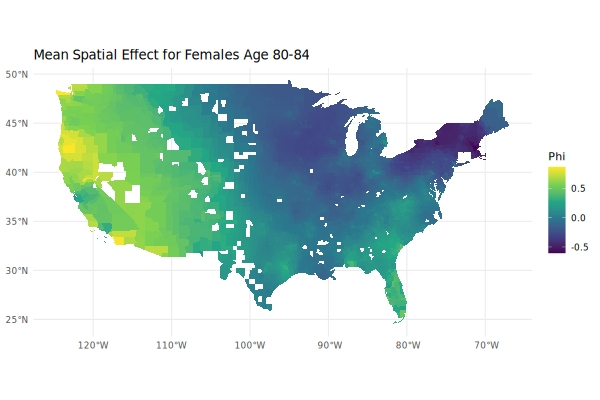}
    \caption{Females, 80--84}
    \label{fig:phi_f_80_84}
  \end{subfigure}

  \vspace{0.6em}

  \begin{subfigure}[t]{0.32\textwidth}
    \centering
    \includegraphics[width=\textwidth]{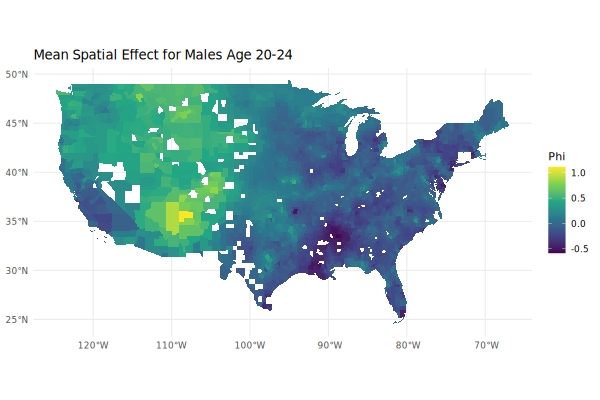}
    \caption{Males, 20--24}
    \label{fig:phi_m_20_24}
  \end{subfigure}%
  \hfill
  \begin{subfigure}[t]{0.32\textwidth}
    \centering
    \includegraphics[width=\textwidth]{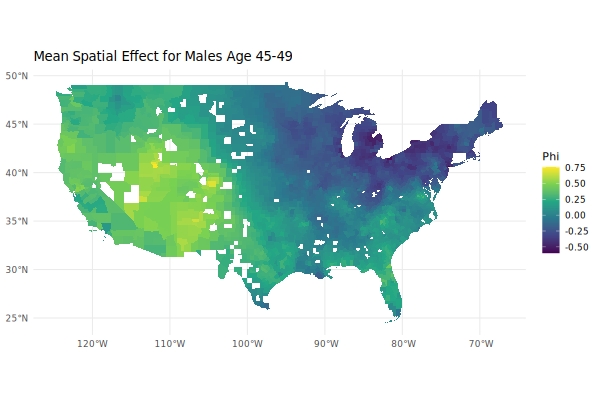}
    \caption{Males, 45--49}
    \label{fig:phi_m_45_49}
  \end{subfigure}%
  \hfill
  \begin{subfigure}[t]{0.32\textwidth}
    \centering
    \includegraphics[width=\textwidth]{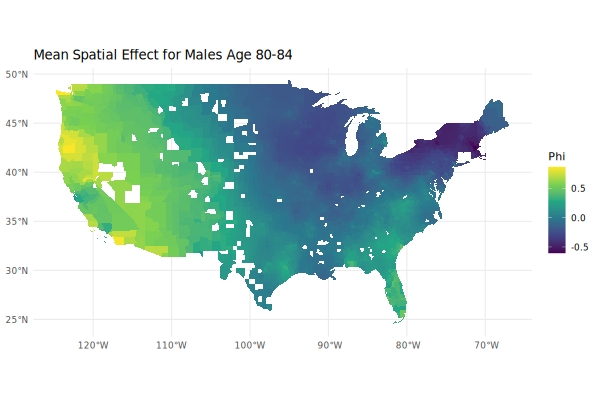}
    \caption{Males, 80--84}
    \label{fig:phi_m_80_84}
  \end{subfigure}

  \caption{Posterior means of the county‐level spatial random effect \(\phi_k\) from the spatio-temporal suicide model.  
  Panels correspond to three age groups (20--24, 45--49, 80--84) and are stratified by sex.  
  Positive values (yellow) indicate counties with higher baseline suicide risk than the national average after adjusting for socio-economic covariates; negative values (blue) indicate lower risk.}
  \label{fig:phi_6panel}
\end{figure}

Figure~\ref{fig:phi_6panel} visualizes the posterior mean of the spatial intercept $\phi_k$, 
which captures county-specific deviations in baseline suicide risk after adjusting for socio-economic 
covariates ($\mathbf{x}_{kt}^\top\boldsymbol{\beta}$) and temporal trends.  
Across all age groups a broad corridor of elevated $\phi_k$ values 
extends from the Rocky Mountain states through the northern Great Plains, a region commonly 
referred to as the suicide belt.  Counties in Utah, Wyoming, Montana, and neighboring areas 
exhibit the largest positive departures, whereas much of the eastern seaboard and parts of the 
Midwest display negative or near--zero effects. These maps highlight enduring 
geographic inequities in baseline suicide risk and underscore the need for region-specific 
strategies.

\begin{figure}[htbp]
  \centering
  \includegraphics[width=0.48\textwidth]{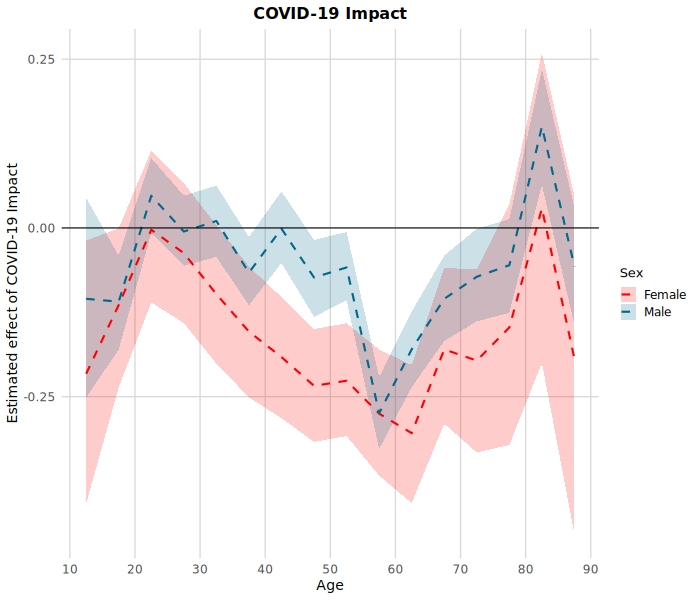}
  \includegraphics[width=0.48\textwidth]{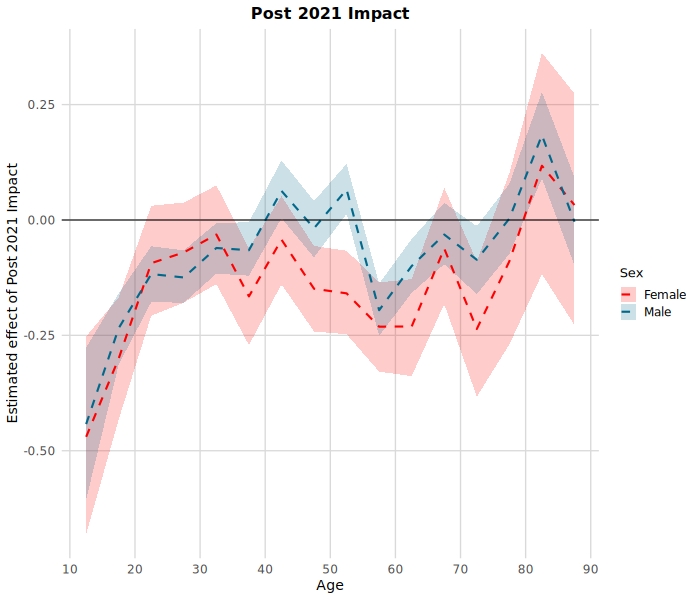}
  \caption{Estimates and 95\% credible intervals for the effect of COVID (2020–2021) and post 2021 (2022–2023) on suicide by sex and age. Negative values indicate reductions relative to the overall time trend.}
  \label{fig:se_covid}
\end{figure}

The COVID variable (indicating years 2020–2021) initially appeared to have a substantial effect, but its magnitude diminished once the “Post 2021” variable (indicating years 2022–2023) was included. Notably, analysis of variable importance indicated that both temporal variables were consistently highly ranked across each age and gender group (see figure \ref{fig:varimportfig} in Appendix \ref{app:varimport}), underscoring their relevance to the models. There is some evidence that suicides decreased during the pandemic, as shown in figure \ref{fig:se_covid}, particularly among women aged 30–70. This may relate to higher overall mortality during the pandemic or to changes in social or environmental factors that benefitted some individuals with pre-existing mental health challenges. While our data do not allow us to draw conclusions regarding what these factors might be, some research indicates, for example, that during times of widespread crisis, individuals and communities pull together in their efforts to manage the catastrophe. In a study of undergraduate students after a significant flood, \cite{gordon2011impact} found that contributing to the crisis response decreased a sense of burdonsomeness and increased a sense of belonging (both of which are predictors of suicide in the interpersonal theory of suicide; \citealp{joiner2010interpersonal}). It is also possible that increased time at home due to initial lockdown measures were a protective factor for some, increasing oversight for those with preexisting suicidality and increasing regular social contact. Supporting this possibility, in a study of 8 million helpline calls from over 19 countries, \cite{brulhart2021mental} found that complaints regarding relationship issues, violence, and suicidal ideation all decreased during the early stages of the pandemic. They additionally found that complaints regarding economic problems decreased, potentially pointing to the ameliorating effects of financial help from the government during the crisis. An increased sense of community purpose, increased home time and social contact, and governmental financial support may at least partially explain the decrease in suicide deaths. The post 2021 effect, however, does not show a consistent relationship with suicide rates.

\begin{figure}[htbp]
  \centering
  \begin{subfigure}[t]{0.48\textwidth}
    \centering
    \includegraphics[width=\textwidth]{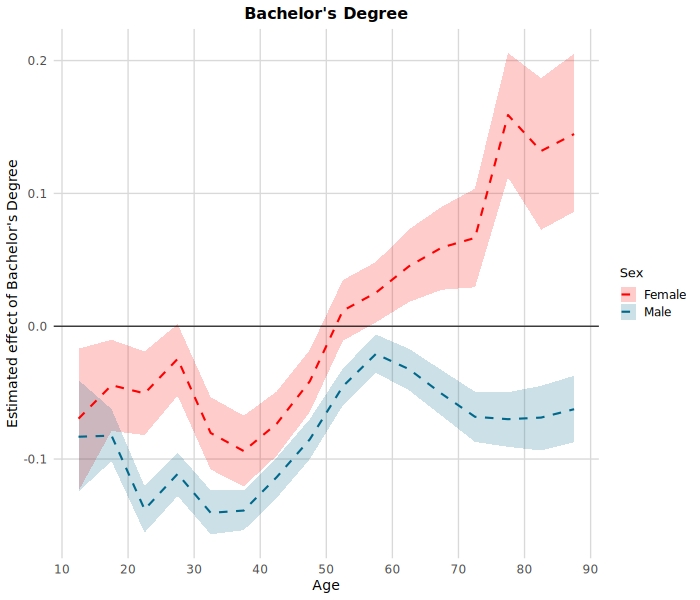}
    \caption{Bachelor’s Degree}
    \label{fig:se_bach}
  \end{subfigure}%
  \hfill
  \begin{subfigure}[t]{0.48\textwidth}
    \centering
    \includegraphics[width=\textwidth]{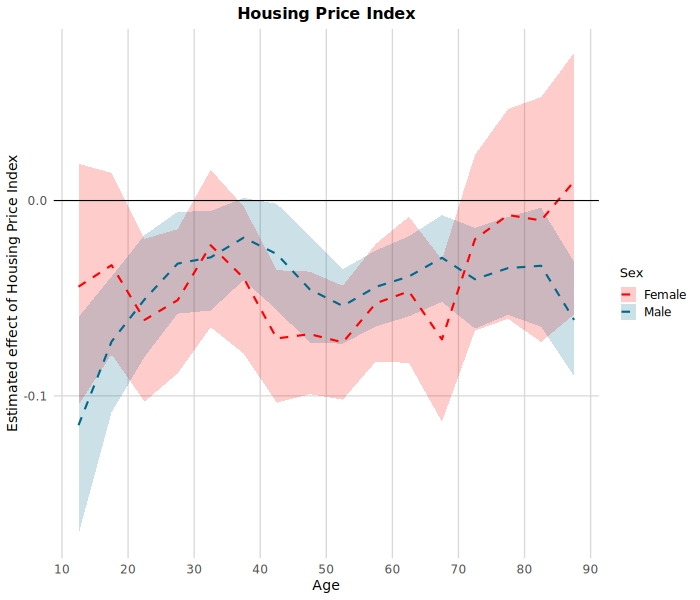}
    \caption{House Price Index (HPI)}
    \label{fig:se_HPI}
  \end{subfigure}

  \vspace{0.5em}
  \begin{subfigure}[t]{0.48\textwidth}
    \centering
    \includegraphics[width=\textwidth]{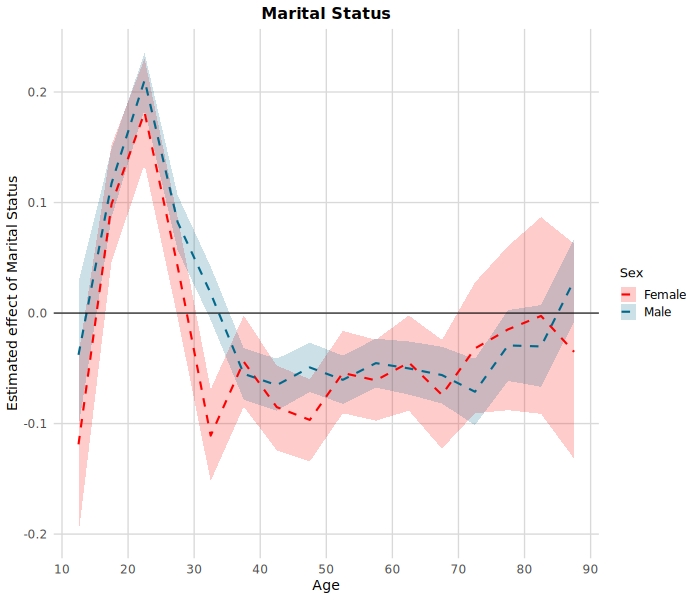}
    \caption{Marriage Rates}
    \label{fig:se_marr}
  \end{subfigure}%
  \hfill
  \begin{subfigure}[t]{0.48\textwidth}
    \centering
    \includegraphics[width=\textwidth]{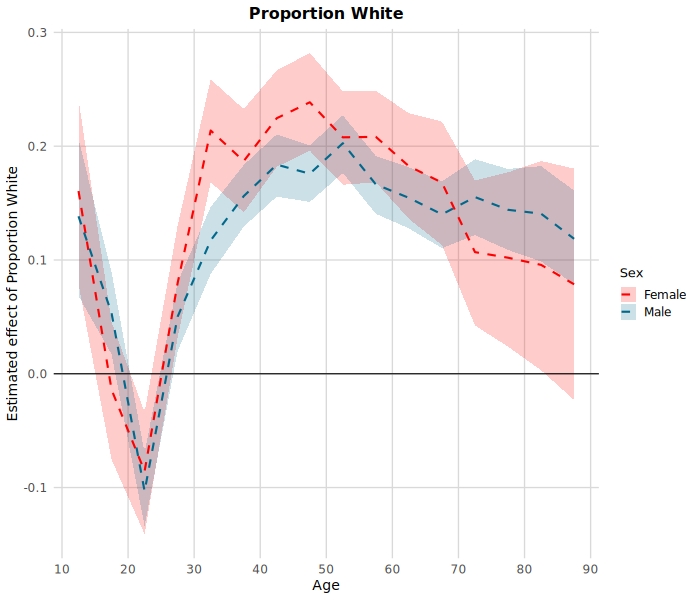}
    \caption{Percent White}
    \label{fig:se_white}
  \end{subfigure}

  \vspace{0.5em}
  \begin{subfigure}[t]{0.48\textwidth}
    \centering
    \includegraphics[width=\textwidth]{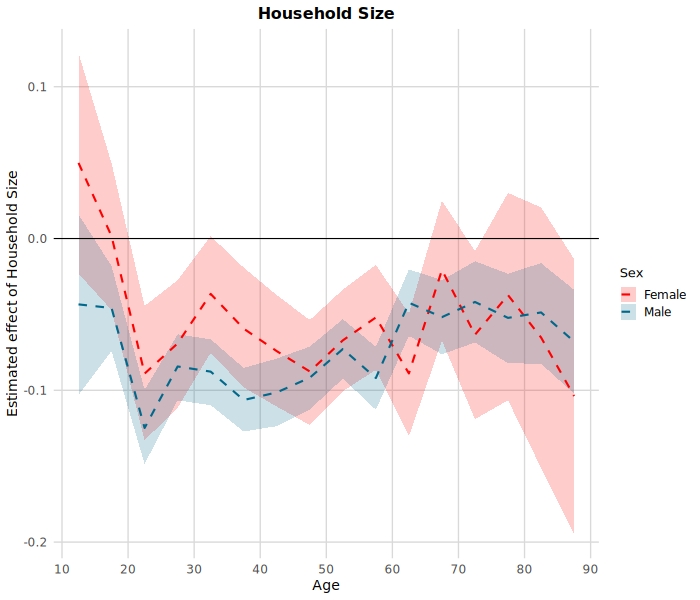}
    \caption{Household Size}
    \label{fig:se_size}
  \end{subfigure}%
  \hfill
  \begin{subfigure}[t]{0.48\textwidth}
    \centering
    \includegraphics[width=\textwidth]{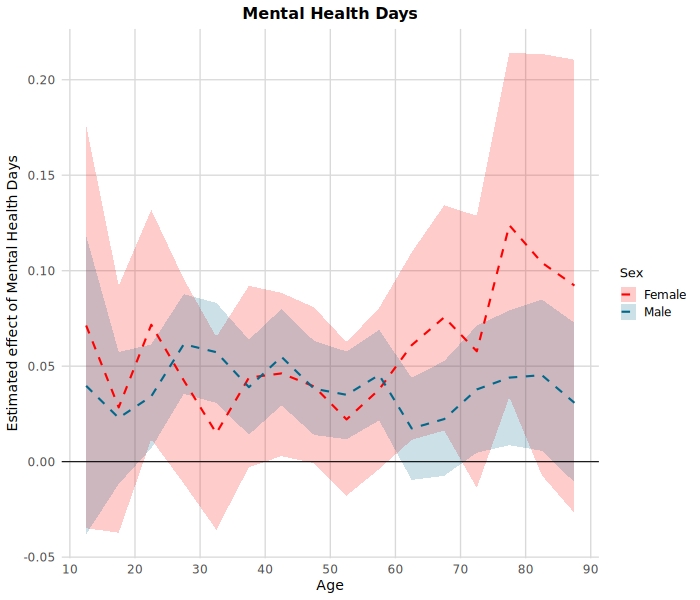}
    \caption{Mental Health Days}
    \label{fig:se_MHD}
  \end{subfigure}

  \caption{Estimates and 95\% credible intervals for six significant covariates on suicide by sex and age.}
  \label{fig:se_sigcovars}
\end{figure}

\subsubsection{Socio-economic Effects}
\label{Socio-economic Effects}

There were nine additional socio-economic or demographic variables included as predictors for each model. Of those, six have interesting patterns worth discussing while three consistently showed little to no effect for all age groups. Figure~\ref{fig:se_sigcovars} summarizes how the six significant covariates
relate to suicide risk across the life-course, separately for men (blue) and
women (red). 

\begin{itemize}
\item \textbf{Educational Attainment (\% with a Bachelor's Degree)} - Among men, a greater county-level share of adults holding a bachelor’s degree is strongly associated with lower suicide rates, with the largest protective association concentrated in early to mid-adulthood (roughly ages 20–40). In contrast, the female pattern reverses across the life course. The association is negative at younger ages but steadily weakens after age 40 and turns positive by about age 50, indicating that in later life, women living in more highly educated counties experience higher suicide rates than their peers elsewhere. Educational attainment therefore represents the clearest point of divergence between the sexes. While most other covariates show only modest male–female differences in magnitude, none display the pronounced change in sign seen for education.
\item \textbf{Housing Price Index} - Across most of the life course, higher county-level housing values are linked to lower suicide rates for both sexes, and the strength of this inverse association remains fairly stable. Among males, the effect becomes even stronger at the age extremes. Adolescents and early adults ($<$ 20 yr) and older ages ($\geq$ 80 yr) have a more negative effect whereas for females the coefficients taper toward zero in those same bands. Outside these margins, male and female estimates are nearly the same.
\item \textbf{Marriage Rates} - Marriage rates have a very interesting trend. For the youngest age group shown (10-14) higher marriage rates are associated with lower suicides, but the effect quickly spikes and the relationship becomes positive starting at age 15. For females, the relationship becomes negative again for ages 30+ while for males it becomes negative at 35+ and it remains significantly negative until about age 70 where the effect for both males and females disappears. 
\item \textbf{Percent White} - The percentage of white heads of households in a county shows almost an inverted effect from marriage rates. This variable is ranked first in importance across all three model groupings (all models, females, and males). For the lowest age group, it is positively associated with suicide rates but then drops to negatively associated between ages 15 to 30, after which the effect becomes strongly positive again. The effect tapers slightly by age but remains significantly positive through older ages. 
\item \textbf{Household size} - The average number of members per household has little to no effect for the youngest age groups, but then becomes significantly negative after about age 20 for both males and females. The effect is nearly constant past age 20 and is about the same for both males and females. 
\item \textbf{Mental Health Days} - The results of the BRFSS on mental health is positively associated with suicide rates. This effect is constant, and while significant, it is actually not far from zero. In fact, the confidence interval for females contains zero for most age groups and does for males at some ages. Starting at age 60 the effect becomes stronger for females than it does for males. 
\end{itemize}

\begin{figure}[htbp]
  \centering
  \includegraphics[width=0.32\textwidth]{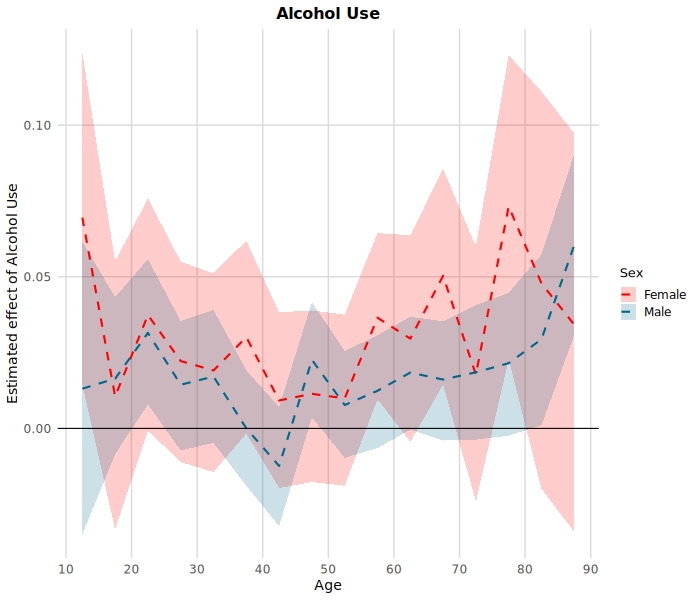}
  \includegraphics[width=0.32\textwidth]{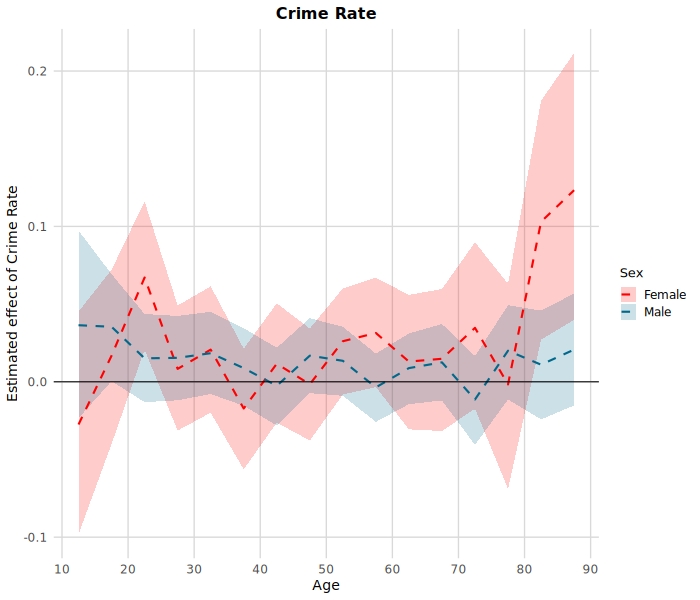}
  \includegraphics[width=0.32\textwidth]{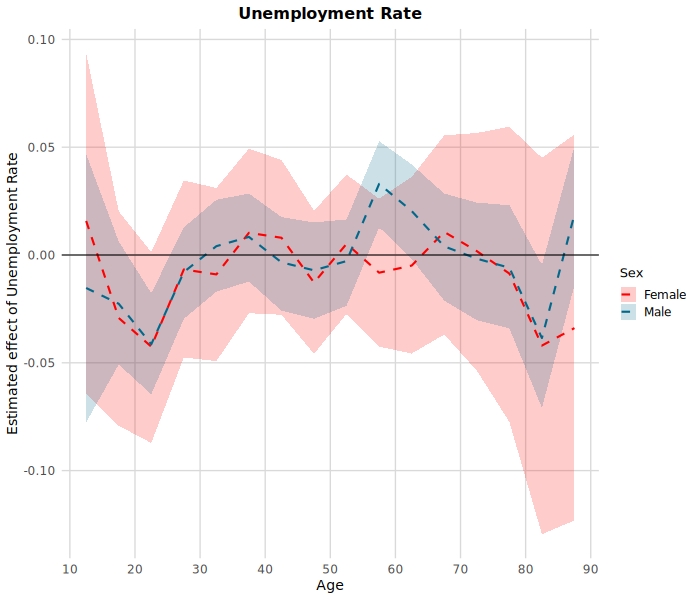}
  \caption{Estimates and 95\% credible intervals for the effect of Alcohol Consumption, Crime Rates, and Unemployment Rates on suicide by sex and age.}
  \label{fig:se_nonsigcovars}
\end{figure}

There were three variables that showed little to no association with suicide rates. Alcohol consumption shows a weak, non-significant tendency toward higher suicide rates across ages.  Neither crime rates nor unemployment rates display any consistent or significant association with suicide rates in this analysis. Figure \ref{fig:se_nonsigcovars} shows these variables, and while there may be some signal or interpretation from specific components of these plots, it is difficult to conclude with certainty that a significant effect exists. 

\subsection{Mental Health Indicators and Mortality Outcomes}

To evaluate how county-level mental health relates to mortality, we present results from models that pair each mental health indicator with two outcomes: all-cause mortality and suicide. This organization directly addresses our second research question by comparing whether associations seen at the population level are broad (all-cause) or more specific to suicide. For each indicator, we show male and female estimates across age groups with 95\% credible intervals.

\subsubsection{Depression and All-Cause Mortality / Suicide}
\label{deathdep}

\begin{figure}[htbp]
  \centering
  \begin{subfigure}[t]{0.48\textwidth}
    \centering
    \includegraphics[width=\textwidth]{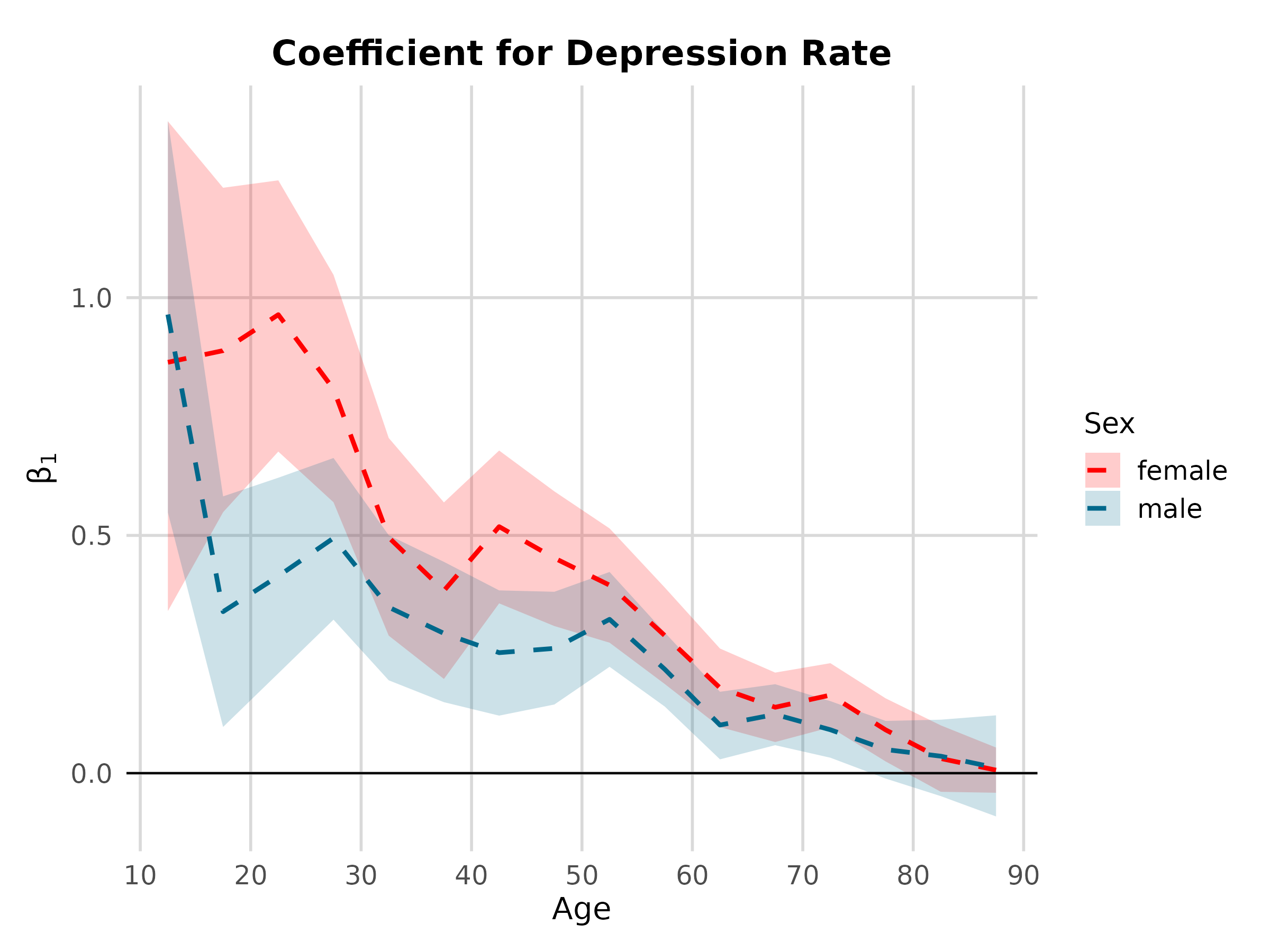}
    \caption{All-cause mortality}
    \label{fig:dep_allcause}
  \end{subfigure}%
  \hfill
  \begin{subfigure}[t]{0.48\textwidth}
    \centering
    \includegraphics[width=\textwidth]{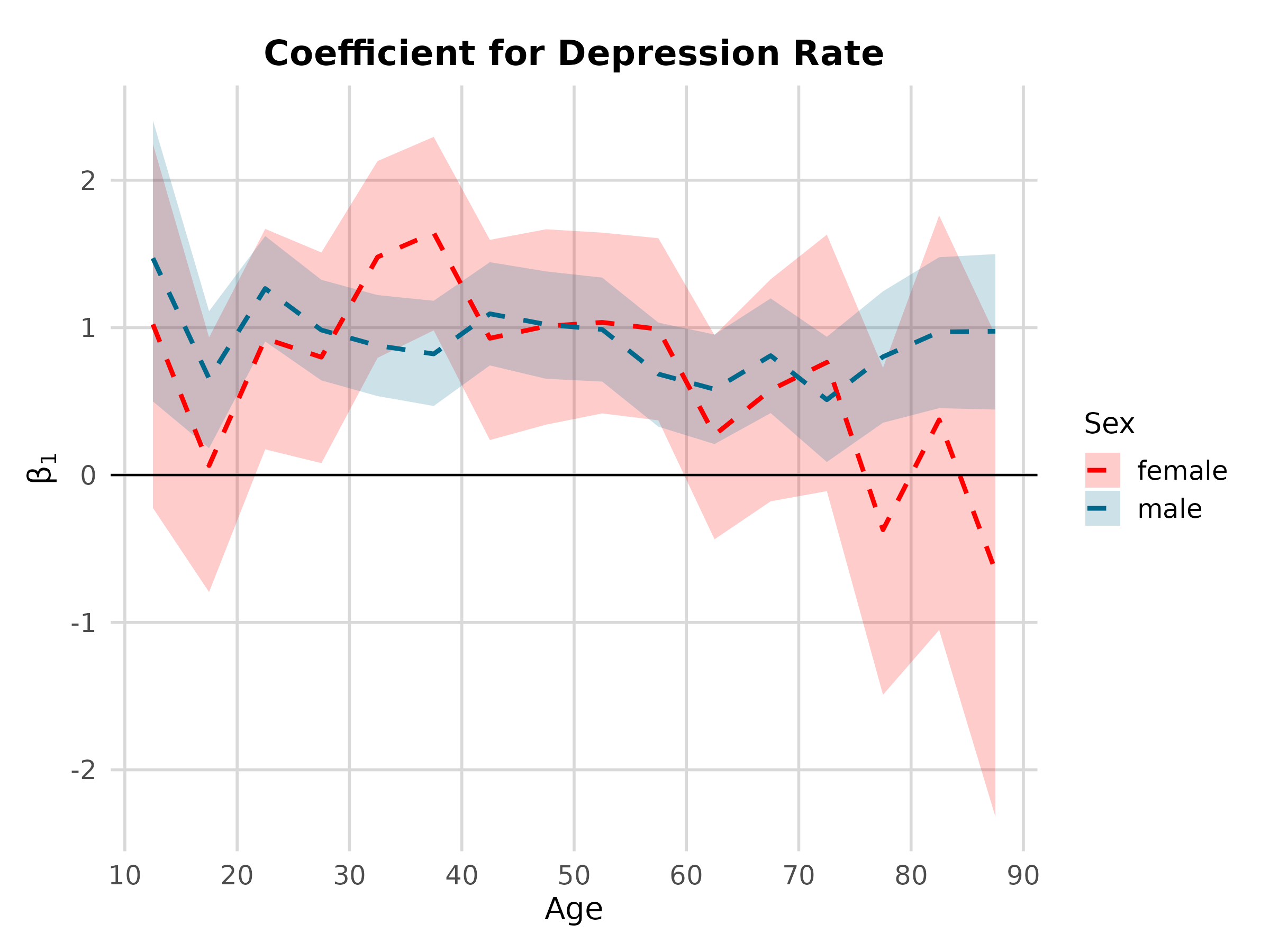}
    \caption{Suicides}
    \label{fig:dep_suicide}
  \end{subfigure}
  \caption{Estimated effect (\(\beta_1\)) of the severe depression positive screening rate on mortality by age group and sex. Shaded bands represent 95\% credible intervals.}
  \label{fig:dep_allcause_vs_suicide}
\end{figure}

Figure \ref{fig:dep_allcause_vs_suicide}\subref{fig:dep_allcause} shows the estimated effect ($\beta_1$) of the severe depression positive screening rate on mortality across age groups, stratified by sex. For both males and females, we observe a positive association in most age groups, suggesting that higher rates of severe depression at the population level are linked with increased mortality risk. The estimated effect is highest among adolescents and young adults, with the largest coefficients observed for individuals under 20. However, the wide credible intervals in these youngest groups reflect greater uncertainty, likely due to lower death counts and smaller sample sizes in these strata.

Across all ages, females tend to exhibit slightly higher effect sizes than males, although there is substantial overlap between the sexes. The magnitude of the association appears to decline steadily with age, becoming close to zero in the oldest age groups. It is important to note that this model estimates mortality at the county-year level, and should not be interpreted as evidence of individual-level causality between severe depression and death. Nonetheless, the population-level association suggests that areas with elevated rates of severe depression may also face higher overall mortality risk, particularly among youth and middle-aged adults.


\subsubsection{Suicidal Ideation and All-Cause Mortality / Suicide}
\label{simort}

\begin{figure}[htbp]
  \centering
  \begin{subfigure}[t]{0.48\textwidth}
    \centering
    \includegraphics[width=\textwidth]{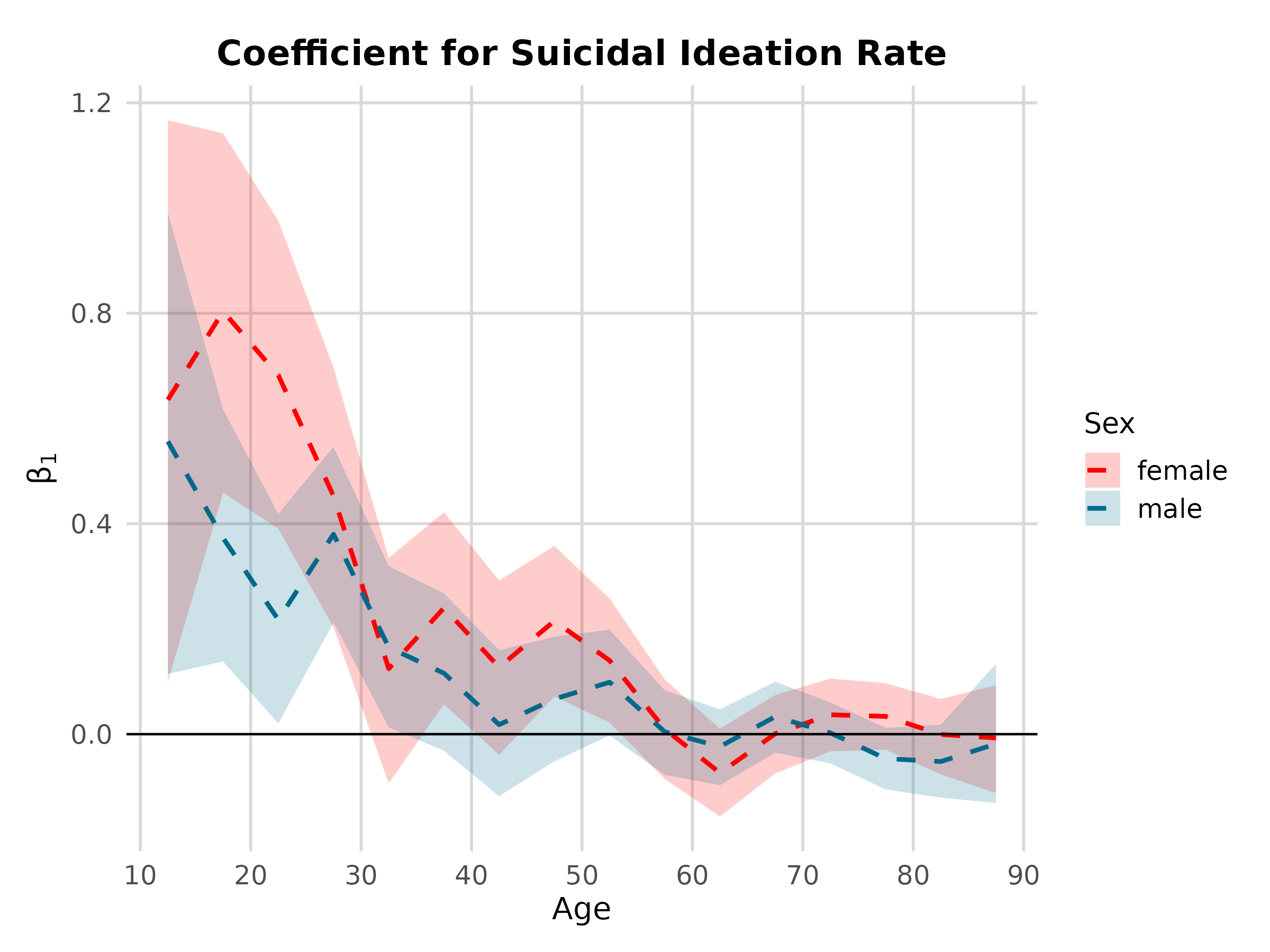}
    \caption{All-cause mortality}
    \label{fig:si_allcause}
  \end{subfigure}%
  \hfill
  \begin{subfigure}[t]{0.48\textwidth}
    \centering
    \includegraphics[width=\textwidth]{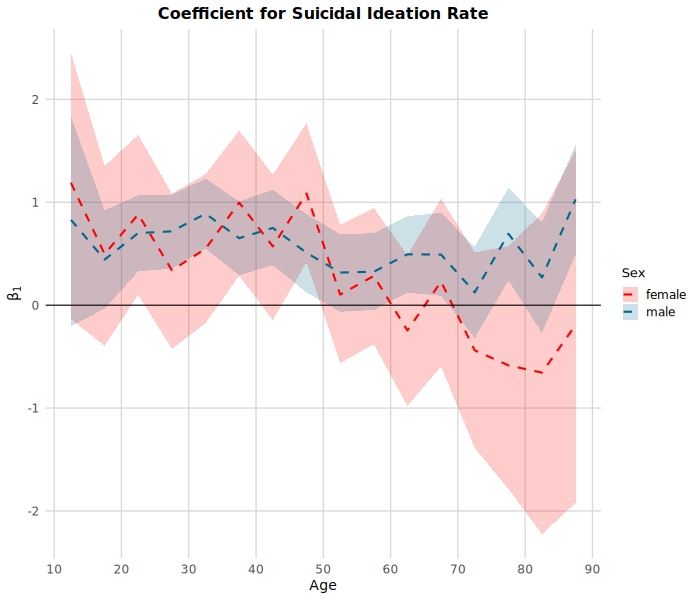}
    \caption{Suicides}
    \label{fig:si_suicide}
  \end{subfigure}
  \caption{Estimated coefficient ($\beta_1$) for the association between the positive screen rate for frequent suicidal ideation and mortality outcomes, stratified by sex and age group. Shaded regions represent 95\% credible intervals.}
  \label{fig:si_allcause_vs_suicide}
\end{figure}

Figure~\ref{fig:si_allcause_vs_suicide}\subref{fig:si_allcause} displays the estimated regression coefficients ($\beta_1$) for the association between the positive screen rate (PSR) for frequent suicidal ideation and mortality rates across age groups, stratified by sex. A positive coefficient indicates that higher rates of suicidal ideation in a county are associated with increased mortality rates in that demographic group.

The association is strongest among younger age groups, particularly for individuals under age 25. For both males and females, the effect size is largest in childhood and adolescence, with credible intervals suggesting a significant positive relationship. Among females in particular, the association peaks in the 10--14 age group before declining steadily with age. After age 40, the estimated coefficients approach zero, and the 95\% credible intervals widen and generally include zero, indicating increasing uncertainty and decreasing evidence of a consistent relationship in older age groups.

These results suggest that frequent suicidal ideation, as measured by community screening, may be a particularly important correlate of mortality risk in younger populations.  The wide intervals in early childhood may also reflect small sample sizes or less stable estimates for those groups.

Figure~\ref{fig:si_allcause_vs_suicide}\subref{fig:si_suicide} displays the estimated effects of the county-level positive screening rate (PSR) for frequent suicidal ideation on suicide counts, stratified by age and sex. In contrast to the model with all-cause mortality as the outcome, this figure directly links suicidal ideation rates with suicide-specific mortality. For both sexes, the coefficients are generally positive in early adulthood and middle ages, with point estimates frequently exceeding 0.5. These results suggest that higher screening rates for suicidal ideation in the population are positively associated with suicide rates.

However, the uncertainty is notable, especially among females. The credible intervals widen substantially in older age groups and cross zero frequently, reflecting limited data and potential variability in the relationship across counties and time. The female estimates decline steeply in older ages, becoming negative in the 70+ range with wide uncertainty, which may stem from small sample sizes or complex age-related reporting dynamics. Male estimates remain more consistent across the lifespan, with point estimates staying near or above 0.5 in most age groups.

These results support the use of community-level mental health screening data as an informative indicator of suicide risk, particularly in younger populations.

\subsubsection{BRFSS Mental Health Days and All-Cause Mortality / Suicide}
\label{deathbrffs}

\begin{figure}[htbp]
  \centering
  \begin{subfigure}[t]{0.48\textwidth}
    \centering
    \includegraphics[width=\textwidth]{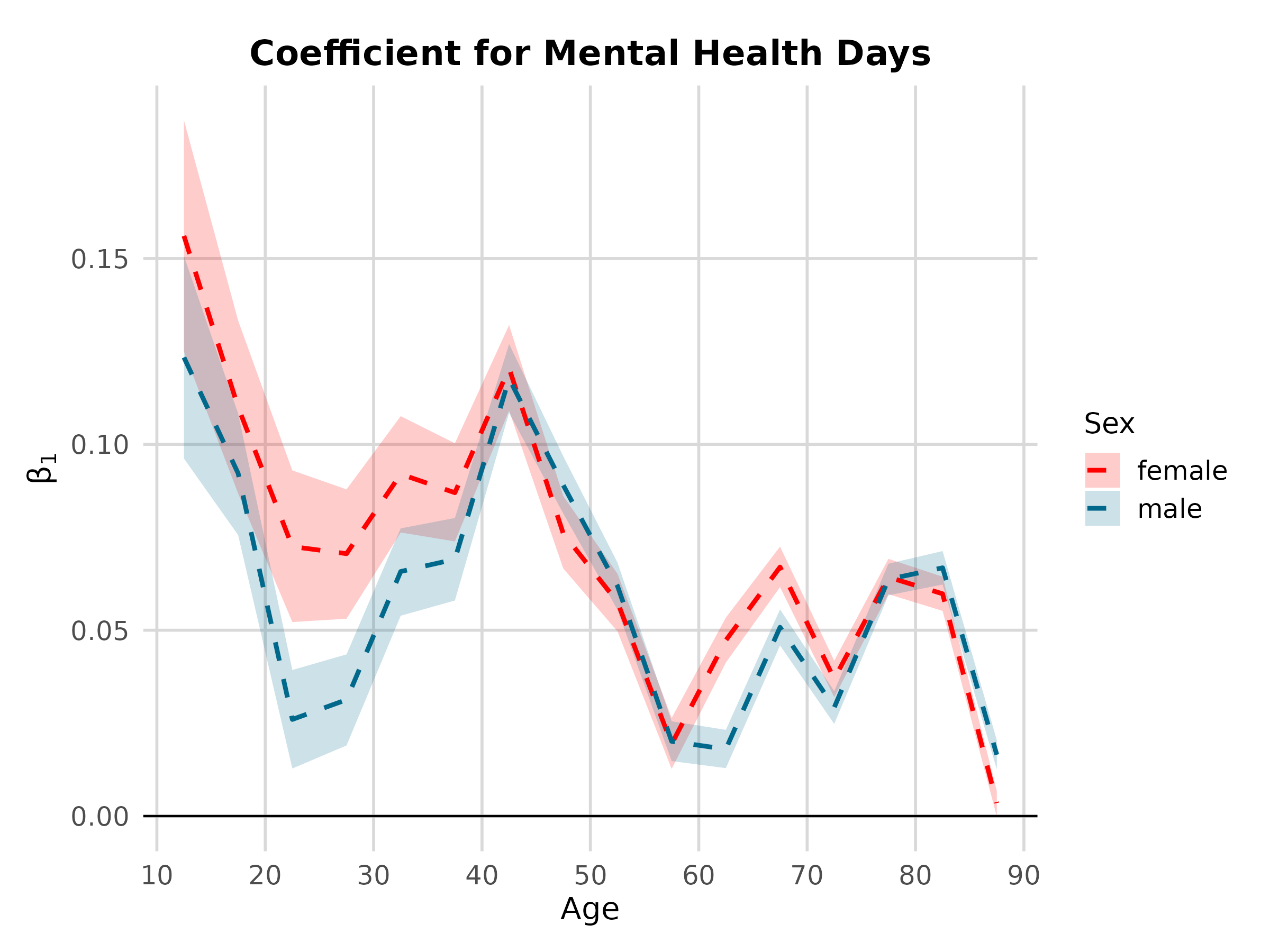}
    \caption{All-cause mortality}
    \label{fig:mhd_allcause}
  \end{subfigure}%
  \hfill
  \begin{subfigure}[t]{0.48\textwidth}
    \centering
    \includegraphics[width=\textwidth]{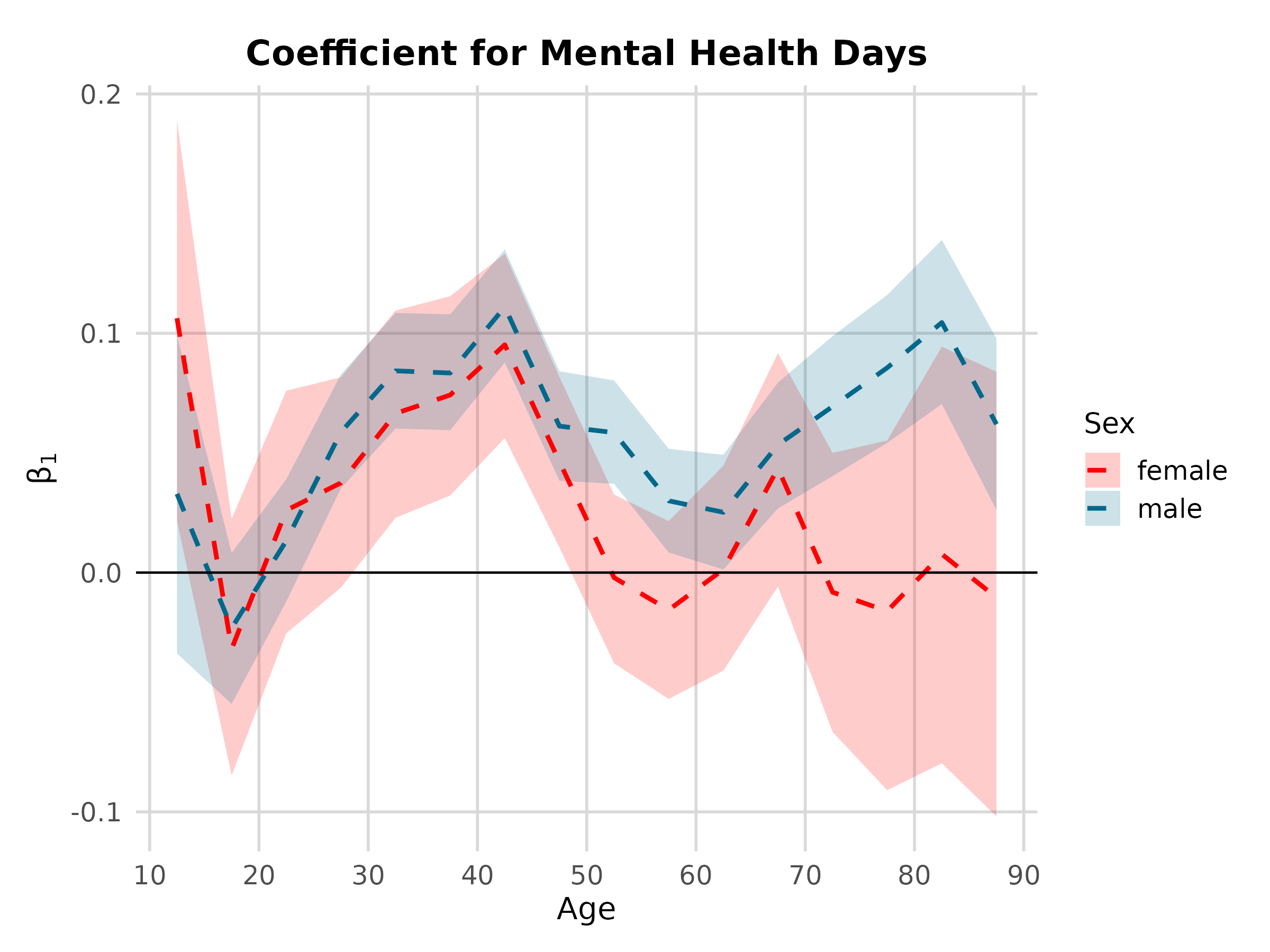}
    \caption{Suicides}
    \label{fig:mhd_suicide}
  \end{subfigure}
  \caption{Estimated coefficient ($\beta_1$) for the association between average number of poor mental health days (past 30 days; BRFSS 2010--2023) and mortality outcomes, stratified by age group and sex. Shaded regions represent 95\% credible intervals.}
  \label{fig:mhd_allcause_vs_suicide}
\end{figure}

Figure~\ref{fig:mhd_allcause_vs_suicide}\subref{fig:mhd_allcause} shows the estimated association between the average number of mentally unhealthy days reported in the past 30 days and mortality rates, using Behavioral Risk Factor Surveillance System (BRFSS) data from 2010 to 2023. As with previous models, estimates are stratified by age group and sex, with shaded regions denoting 95\% credible intervals.

The results suggest a modest but consistently positive association between reported mental health burden and mortality, especially in younger age groups. For both sexes, the association peaks in early adolescence (age group 10--14) and declines steadily through midlife. This pattern aligns with previous findings indicating that self-reported mental distress is more prevalent and perhaps more impactful on broader health outcomes during adolescence and early adulthood.

Interestingly, both males and females show a secondary peak in association around the 45--49 age group, particularly in males, before declining again in later life. In the oldest age groups (70+), the estimated effects are smaller and less certain, as indicated by wider credible intervals.

These findings support the use of the BRFSS mental health measures as a meaningful covariate in population-level mortality modeling.

Figure~\ref{fig:mhd_allcause_vs_suicide}\subref{fig:mhd_suicide} displays the estimated relationship between the average number of poor mental health days (reported in the past 30 days) and suicide counts across U.S. counties. Unlike previous models focused on all-cause mortality, this model isolates suicide as the outcome of interest, providing a more targeted view of how self-reported mental health burden correlates with suicide risk.

Overall, the estimated associations are smaller in magnitude compared to those seen in the mortality model, and they demonstrate more variability across age groups. For both sexes, coefficients are generally positive through midlife, suggesting that higher rates of poor mental health days in the population are associated with higher suicide rates. However, many of these associations are not statistically distinguishable from zero, as reflected in the wide credible intervals, especially in younger and older age groups.

Among males, the association becomes more pronounced in later age groups (ages 60+), while the trend for females appears relatively flat or slightly negative in older adulthood. This divergence may point to gendered differences in how mental health distress manifests in suicide outcomes later in life. Additionally, the wide intervals highlight uncertainty in these estimates, which may be due in part to smaller suicide counts relative to all-cause deaths and greater year-to-year variation.

As with all ecological models, these results reflect population-level associations and should not be interpreted as direct individual-level causal effects.

\begin{table}[!ht]
\centering
\caption{Covariates and coefficient meanings across model families. Note that fixed effects and random effects estimated separately for each sex $\times$ age group.}
\label{tab:covariate-map}
\renewcommand{\arraystretch}{1.1}
\arrayrulecolor{black}
\begin{tabular}{p{5.0cm} p{2.6cm} p{8.2cm}}
\hline
\textbf{Variable (county-year)} & \textbf{Symbol in text} & \textbf{Coefficient meaning (on log-odds scale)} \\
\hline
\multicolumn{3}{l}{\emph{Socio-economic model of suicide} (Section 4.2.1; 2010--2023; includes COVID indicators)}\\
\hline
Alcohol consumption (state per-capita) & $\beta_{\text{Alcohol}}$ & Change in log-odds per 1-unit increase in alcohol measure. \\
\% Bachelor’s degree (ACS) & $\beta_{\text{Educ}}$ & Change in log-odds per 1 p.p.\ increase in adults with BA+. \\
Crime rate & $\beta_{\text{Crime}}$ & Change in log-odds per 1-unit increase in crime index/rate. \\
House Price Index (FHFA) & $\beta_{\text{HPI}}$ & Change in log-odds per 1-unit increase in HPI. \\
Marriage rate (ACS) & $\beta_{\text{Married}}$ & Change in log-odds per 1 p.p.\ increase in married adults. \\
Household size (ACS) & $\beta_{\text{HHSize}}$ & Change in log-odds per 1-person increase in mean household size. \\
Unemployment rate (BLS) & $\beta_{\text{Unemp}}$ & Change in log-odds per 1 p.p.\ increase in unemployment. \\
\% White heads of household (ACS) & $\beta_{\text{Race}}$ & Change in log-odds per 1 p.p.\ increase in white HoH. \\
Poor mental health $\ge 14$ days (BRFSS/PLACES) & $\beta_{\text{MHdays}}$ & Change in log-odds per 1 p.p.\ increase in frequent mental distress. \\
COVID period indicator (2020--2021) & $\beta_{\text{COVID}}$ & Shift in log-odds during 2020--2021 vs.\ pre-2020. \\
Post 2021 indicator (2022--2023) & $\beta_{\text{Post 2021}}$ & Shift in log-odds during 2022--2023 vs.\ pre-2020. \\
\hline
\multicolumn{3}{l}{\emph{Mental health models} (Section 4.2.2; each MH indicator paired with two outcomes: all-cause mortality and suicide deaths)}\\
\hline
Severe depression positive screen rate (MHA; PHQ-9) & $\beta_{\text{PSR,Dep}}$ & Change in log-odds per 1 p.p.\ increase in PSR (severe depression). \\
\arrayrulecolor{gray!60}\hline
Frequent suicidal ideation PSR (MHA; PHQ-9 item 9) & $\beta_{\text{PSR,SI}}$ & Change in log-odds per 1 p.p.\ increase in PSR (SI). \\
\arrayrulecolor{gray!60}\hline
PTSD PSR (MHA; PC-PTSD-5) & $\beta_{\text{PSR,PTSD}}$ & Change in log-odds per 1 p.p.\ increase in PSR (PTSD). \\
\arrayrulecolor{gray!60}\hline
Psychosis-risk PSR (MHA; PQ-B) & $\beta_{\text{PSR,Psych}}$ & Change in log-odds per 1 p.p.\ increase in PSR (psychosis risk). \\
\arrayrulecolor{gray!60}\hline
Poor mental health days (BRFSS/PLACES) & $\beta_{\text{MHdays}}$ & Change in log-odds per 1 p.p.\ increase in frequent mental distress. \\
\arrayrulecolor{black}\hline
\end{tabular}
\end{table}

\section{Conclusion}
\label{sec:conc}

We have examined the relationships between mental health status and overall mortality and suicides at the county level in the continental U.S. We first assessed socio-economic covariates in predicting suicide. While the results vary considerably by age and sex, we found that the county-wide levels of educational attainment, housing prices, marriage rates, racial composition, household size, and poor mental health days all have significant relationships with suicide rates. To further contextualize these findings, Appendix \ref{app:varimport} presents the ranked importance of each variable across all fitted models, highlighting the relative influence of demographic, economic, and behavioral factors on county-level suicide risk.

We also examined the impact of various mental health indicators on all-cause and suicide-specific mortality and found that the strongest effects are observed in younger populations. The spatial and temporal correlation structures revealed substantial regional clustering and time-consistent trends in both all-cause mortality and suicide rates, supporting the use of spatio-temporal methods. Our findings highlight the value of integrating mental health surveillance data into mortality models to better identify emerging risk areas and vulnerable populations. This approach has the potential to inform public health policy, resource allocation, and targeted interventions aimed at reducing disparities in mortality and suicide across U.S. communities.

There are many directions in which this work might be extended in the future. As noted above, our covariates are measured at the county level. Thus, while they are helpful in indicating associations, it is difficult to use them to make individual-level causal statements; more granular data might help to clarify some of these relationships. It is also important to note that some of our mental health data only comprises a few years of information, making it difficult to disentangle any COVID (and post 2021) effects. Finally, while our model is able to capture both spatial and temporal effects, more flexible models might be able to discover more complex patterns in the data.

\section*{Acknowledgments}

We thank the members of Project Oversight Group and staff from the Society of Actuaries Research Institute (namely Tamara Bogojevic-Catanzano, Bryan Burningham, Kara Clark, Jean-Marc Fix, Rachelle Jacobs, Maggie Ma, Murali Niverthi, Marianne Purushotham, Becca Reppert, Barbara Scott, and Zhanxiong Song) whose helpful suggestions, guidance, and comments have served to improve the quality of this project. We also thank Ben Lander, who added valuable insights into the data. Finally, we are grateful for the support of the Society of Actuaries Research Institute, the BYU Department of Statistics, and Robert Morris University, which funded the project.

\bibliographystyle{apalike}
\bibliography{references}
\newpage
\begin{appendices}

\section{Data Adjustments}
\label{app:dataadjust}

\begin{table}[htbp]
\centering
\caption{Mapping of Original to Adjusted FIPS Codes. The adjusted FIPS represents the county code after merging or reallocation due to population considerations.}
\label{tab:fips_changes}
\begin{tabular}{|cc|p{0.2 in}|cc|p{0.2 in}|cc|}
\hline
\textbf{Original} & \textbf{Adjusted} & & \textbf{Original} & \textbf{Adjusted}  & & \textbf{Original} & \textbf{Adjusted} \\
\hline
01011 & 01101 && 21105 & 21083 && 31009 & 31041 \\
05037 & 05035 && 21129 & 21065 && 31057 & 31029 \\
06003 & 06017 && 21165 & 21173 && 31075 & 31031 \\
08017 & 08063 && 21197 & 21049 && 31085 & 31111 \\
08023 & 08003 && 21237 & 21175 && 31103 & 31089 \\
08049 & 08069 && 22013 & 22015 && 31105 & 31033 \\
08053 & 08067 && 22035 & 22083 && 31113 & 31111 \\
08057 & 08069 && 22037 & 22033 && 31115 & 31041 \\
08061 & 08089 && 22059 & 22079 && 31117 & 31111 \\
08079 & 08007 && 22091 & 22033 && 31171 & 31031 \\
08093 & 08059 && 22107 & 22041 && 32009 & 32023 \\
08111 & 08067 && 28023 & 28075 && 32011 & 32007 \\
13007 & 13095 && 28055 & 28151 && 32015 & 32007 \\
13035 & 13151 && 28063 & 28085 && 32017 & 32003 \\
13037 & 13095 && 28069 & 28075 && 32021 & 32019 \\
13053 & 13215 && 28097 & 28043 && 32027 & 32031 \\
13181 & 13073 && 28125 & 28151 && 32029 & 32031 \\
13271 & 13069 && 28143 & 28033 && 35011 & 35005 \\
13307 & 13261 && 28163 & 28049 && 35033 & 35049 \\
16025 & 16039 && 29103 & 29001 && 37095 & 37013 \\
16033 & 16051 && 30007 & 30031 && 37103 & 37133 \\
16081 & 16019 && 30025 & 30017 && 37177 & 37055 \\
17069 & 17165 && 30055 & 30085 && 38007 & 38089 \\
20019 & 20035 && 30069 & 30027 && 40057 & 40065 \\
20081 & 20055 && 30107 & 30027 && 46017 & 46015 \\
21005 & 21073 && 30109 & 30083 && 46041 & 46129 \\
21063 & 21043 && 31005 & 31101 && 46063 & 46019 \\ \hline
\end{tabular}
\end{table}

\newpage
\section{Variable Importance Rankings}
\label{app:varimport}

\begin{figure}[htbp]
  \centering
  \includegraphics[width=0.32\textwidth]{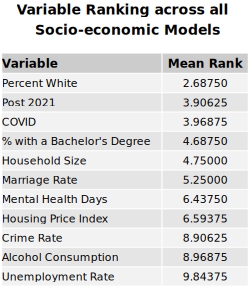}
  \includegraphics[width=0.32\textwidth]{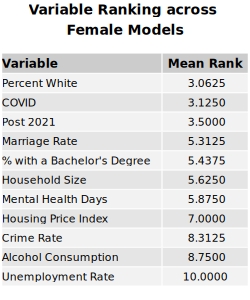}
  \includegraphics[width=0.32\textwidth]{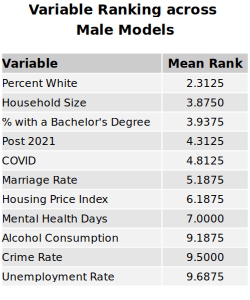}
  \caption{Mean ranks of socio-economic predictors across 32 age–gender suicide models (ages 10–14 to 85+). Variables are ordered by the absolute magnitude of their effects. The table on the left includes all 32 models, and the middle and right tables show results for only females and only males, respectively.}
  \label{fig:varimportfig}
\end{figure}

\newpage
\section{Supplemental Plots}
\label{app:suppplots}
\vspace{0.5 in}
\begin{figure}[htbp]
  \centering
  \begin{subfigure}[t]{0.48\textwidth}
    \centering
    \includegraphics[width=\textwidth]{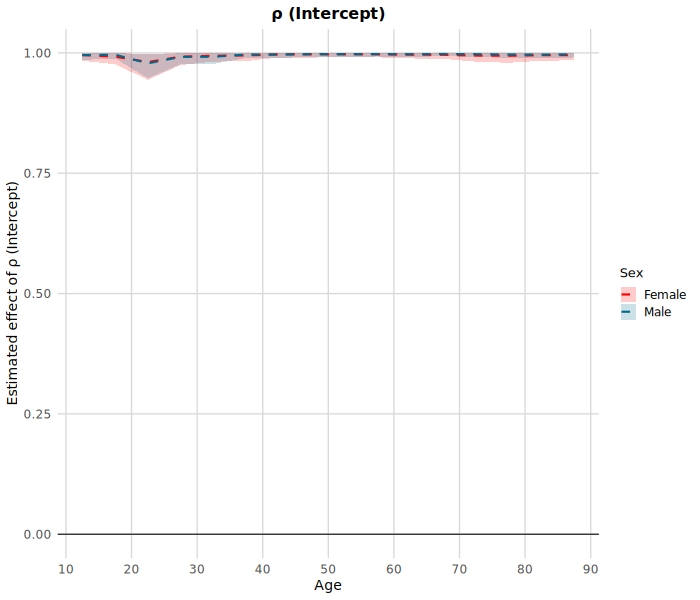}
    \caption{Spatial dependence parameter by age and sex $(\rho_\text{int}$).}
    \label{fig:se_rho_int}
  \end{subfigure}%
  \hfill
  \begin{subfigure}[t]{0.48\textwidth}
    \centering
    \includegraphics[width=\textwidth]{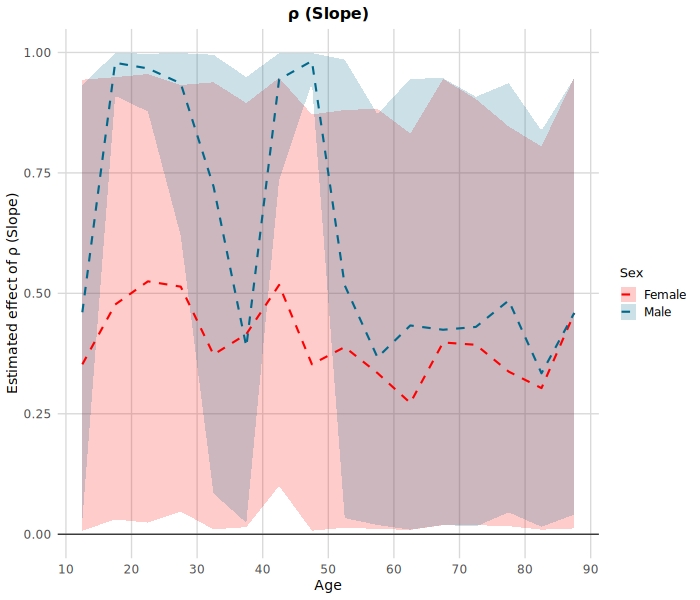}
    \caption{Temporal dependence parameter by age and sex $(\rho_\text{slo})$.}
    \label{fig:se_rho_slo}
  \end{subfigure}
  
  \caption{Estimates and 95\% credible intervals for spatial (\(\rho_\text{int}\)) and temporal (\(\rho_\text{slo}\)) dependence parameters by age and sex.}
  \label{fig:rho_params_side_by_side}
\end{figure}

  \begin{figure}[htbp]
    \centering
    \includegraphics[width=0.8\textwidth]{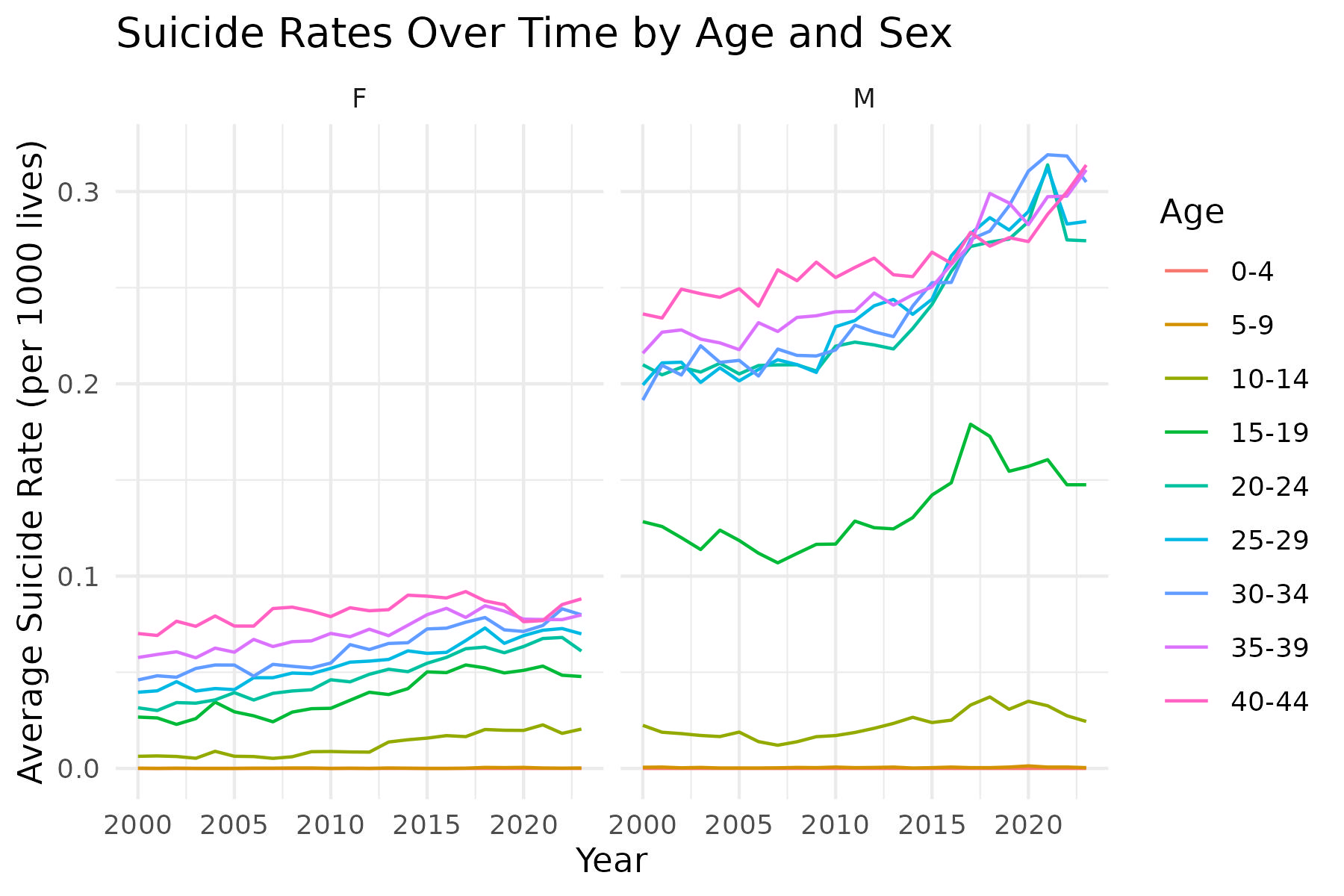}
    \caption{Trends in age- and sex-specific suicide rates in the United States, 2000--2023, based on mortality data from the National Center for Health Statistics. This graph shows the average suicide rates over time for individuals under age 45. Rates are expressed as deaths per person-year within each age-sex group.}
    \label{fig:suicide_rates_45_down}
\end{figure}

  \begin{figure}[htbp]
    \centering
    \includegraphics[width=0.8\textwidth]{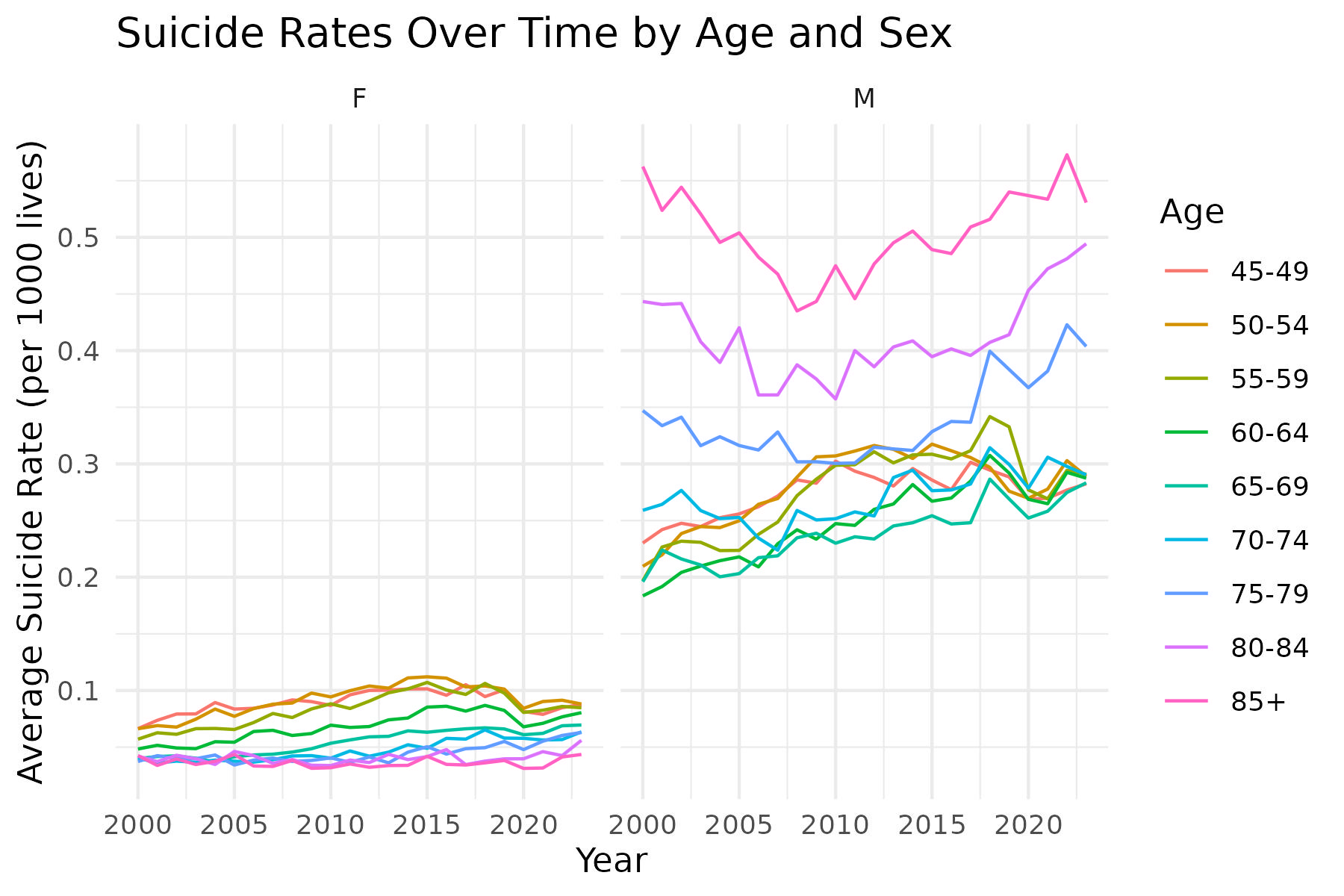}
    \caption{Trends in age- and sex-specific suicide rates in the United States, 2000--2023, based on mortality data from the National Center for Health Statistics. This graph shows the average suicide rates over time for individuals aged 45 and older. Rates are expressed as deaths per person-year within each age-sex group.}
    \label{fig:suicide_rates_45_up}
\end{figure}

\end{appendices}
\end{document}